\documentstyle[11pt,aaspp4,flushrt,epsfig]{article}

\def\deg{$^{\circ}$}

\lefthead{Slee et al.}
\righthead{Relic Radio Sources in Clusters}

\begin{document}

\title{Four Extreme Relic Radio Sources in Clusters of Galaxies}

\author{O. B. Slee}
\affil{Australia Telescope National Facility, CSIRO, PO Box 76, Epping, NSW
1710, Australia}  

\author{A. L. Roy}
\affil{Max-Planck-Institut f\"ur Radioastronomie, Auf dem H\"ugel 69,
    D--53121 Bonn, Germany}

\author{M. Murgia}
\affil{$^1$~Istituto di Radioastronomia del CNR, Via Gobetti 101, 
I--40129 Bologna, Italy\linebreak[4]
$^2$~Dipartimento di Fisica, Universit\`a di Bologna, 
Via B. Pichat 6/2, I--40127 Bologna, Italy}

\author{H. Andernach}
\affil{Depto.\ de Astronom\'\i a, Univ.\ Guanajuato, Apdo.\ Postal 144, 
Guanajuato, C.P.\ 36000, GTO, Mexico}

\and

\author{M. Ehle}
\affil{$^1$~XMM-Newton Science Operations Centre, Apartado 50727, E--28080 
Madrid, Spain\linebreak[4]
$^2$~Astrophysics Division, Space Science Department of ESA, ESTEC,
2200 AG Noordwijk, The Netherlands}

\begin{abstract}
We describe the results of the highest-resolution radio observations
yet made of four relic radio sources in the Abell clusters A13, A85,
A133 and A4038. Our VLA images at 1.4\,GHz with 4$''$
resolution and a noise level of 1 $\sigma\sim$
20\,$\mu$Jy\,beam$^{-1}$ (1.1\,K) show a remarkable variety of fine
structure in the form of spectacular arcs, wisps, plumes and loops.
Their integrated radio flux densities fall very rapidly with
frequency, with power-law slopes, $\alpha$, between 2.1 and 4.4 near
1.4\,GHz (where $S_\nu \propto (\nu/\nu_0)^{-\alpha}$).  The relics
possess linear polarization levels ranging between 2.3\,\% (A133) and
35\,\% (A85); the higher polarization fractions imply a highly ordered
magnetic field in the fine structure and low differential Faraday
rotation in the intervening cluster gas.

The optical identification of a host galaxy formerly associated with a
relic remains problematic. In A85, A133 and A4038 the travel times for
the brightest cluster galaxies are significantly longer than the modeled 
ages of the relics; there is always at least one nearby relatively bright 
elliptical that provides a better match.

Excess X-ray emission in the 0.5\,keV-to-2\,keV band was found near
the relics in A85 and A133.  The surface brightness was too high to be 
attributed to the inverse-Compton mechanism alone.

We found excellent fits to the broad-band radio spectra using the
anisotropic (KGKP) model of spectral ageing, and we have extended the
model to include diffusion of particles between regions of different
field strength (the Murgia-JP, or MJP, model). The steep radio spectra 
imply ages
for the relics of $\sim 10^8$ yr, at the start of which period their
radio luminosities would have been $\sim 10^{25}$ W Hz$^{-1}$ at 1.4
GHz, and so their progenitors were on the boundary between 
FR\,I and FR\,II radio galaxies and hence among the most luminous
7\,\% of radio galaxies.

We find that the relics are in
approximate pressure equilibrium with the surrounding intra-cluster
gas, which has probably limited their free expansion and prevented
them from fading by adiabatic cooling.
 
\end{abstract}


\section{Introduction} \label{intro}

About one in ten rich clusters of galaxies observed with comparatively
low-resolution radio telescopes contains a diffuse steep-spectrum
source with no optical identification (Giovannini, Tordi \& Feretti
1999). Some of these sources take the form of a halo near the cluster
centre with a diameter of 1\,Mpc to 2\,Mpc; the prototype is the halo
in the Coma cluster (Willson 1970). Others with various morphologies
(sometimes arc-like or with multiple components) are situated further
from the cluster centre (Slee \& Reynolds 1984) or even out near the
cluster perimeter (R\"ottgering et al.\ 1997); such non-central
sources are termed ``relics''.  In the clusters Coma (En{\ss}lin et
al.\ 1998), A1300 (Reid et al.\ 1999), A2255 (Feretti et al. 1997),
A2744 (Govoni et al. 2001) and possibly in the cluster 1E~0657$-$56
(Liang et al. 2000), both a halo and a relic have been found.

Recently, Slee \& Roy (1998) found much fine structure at 
4$''$ resolution in the relic in A4038.  This relic is most
likely the remnant of a radio galaxy whose progenitor is now some
18\,kpc to the relic's east, though surprisingly the structure 
does not look like a relaxed plasma in which no further electron
reacceleration is taking place.

Various schemes for the production of halos and relics include the
remnants of de-energised radio galaxies, adiabatic recompression of old,
expanded radio lobes (En{\ss}lin \& Gopal-Krishna 2001), the turbulent wakes of
galaxies that once passed through the region, reacceleration of
electrons in shock fronts caused by cluster mergers, and
reacceleration of electrons in shocks caused by the accretion of gas
by the cluster potential. Most of these proposals possess the common
feature that adiabatic expansion of the radio plasma has been
prevented by the pressure of the hot intra-cluster gas for
$\sim$\,10$^8$\,yr, allowing the inverse-Compton and synchrotron
losses to steepen the electron-energy spectrum and thus the radio
spectrum (Goldshmidt \& Rephaeli 1994).  Detailed modeling of the
spectral evolution can then yield estimates of the relic's age and
magnetic field strength and, if a direct detection of the 
inverse-Compton X-rays is made, of the magnetic field strength
independent of the usual equipartition arguments.

This paper describes the results of further higher-resolution
observations with the Very Large Array (VLA) of the relics
A13\_6a/b/c, A85\_25a/b/c and A133\_7a/b.  These relics were first
identified in the VLA and Molonglo Observatory Synthesis Telescope 
(MOST) maps of Slee \& Reynolds (1984), who
demonstrated their extremely steep ($\alpha\ge\,$2) spectra and lack
of optical host galaxies (where S$_{\nu} \propto
(\nu/\nu_{\circ})^{-\alpha}$).  Better VLA images were presented by
Slee, Roy \& Savage (1994) of these and other relics that were found
during their survey of 58 clusters drawn from the Abell cluster sample
(Abell, Corwin~Jr., \& Olowin 1989). Here, we present new images at
4$''$ resolution of three of these relics, which is the highest
resolution yet achieved in imaging such sources, and give further
measurements from the images of A4038\_9 (Slee \& Roy 1998).  Basic
properties of the clusters are given in Table~1, where the
nomenclature is that of Slee, Roy \& Savage (1994).

Throughout this paper we adopt $H_{\circ}$ = 75\,km\,s$^{-1}$\,Mpc$^{-1}$ 
and $q_{\circ}$ = 0.5.

\section{The Observations} \label{theobs}

We observed the four relics with the VLA at 1.425\,GHz in the BnA
configuration (A13, A133, A4038) or the B configuration (A85) for 8~h
and 9~h, respectively. A13 and A133 were observed on 1998 June 25, A85
on 1998 September 1 and A4038 on 1997 January 26. We obtained higher
resolution (4$''$) than that of our earlier survey observations
(15$''$) (Slee, Roy \& Savage 1994).  We used four 50 MHz bands,
recording left- and right- and cross-circular polarization products.
Calibration and imaging were done with AIPS, using standard
methods. The flux-density scale was calibrated assuming that 3C~48 had
flux densities of 16.36\,Jy at 1.385\,GHz and 15.67\,Jy at
1.465\,GHz. The polarization calibrators were 3C~286 for A13 and A133
and 3C~138 for A85 and A4038. Nearby phase calibrators were observed
every 35~min throughout the observation for initial phase
calibration. The data were then self-calibrated and separate images
were made for the two frequency bands using uniform and natural
weighting. The dirty images were deconvolved using the CLEAN algorithm,
with component subtraction from the ungridded ($u$,\,$v$) data as
implemented in the AIPS task MX, and the images were restored using
beams of $\sim$\,4$''$ and $\sim$\,6$''$ for the uniformly- and
naturally-weighted images, respectively.  The $1\,\sigma$ noise in the
final images ranged from 17\,$\mu$Jy\,beam$^{-1}$ to 21\,$\mu$Jy\,beam$^{-1}$
in a 4$''$ beam (1.0 K to 1.2\,K), which is $\sim$\,50\,\% above the thermal
noise limit.
 
To measure the spectral index, we used flux densities from  the 1.385
and 1.465\,GHz images which, although closely spaced in frequency, yielded
useful values of spectral index for these steep spectrum sources.
 
The ($u$,\,$v$) data were polarization-calibrated, and polarized intensity
images were made from the resulting images of Stokes parameters Q and U.
The instrumental polarization varies with time, limiting the minimum
believable polarization fraction to nominally 0.15\% for the VLA.

\section{Results}  \label{results}

\subsection{Radio Images}  \label{images}

\begin{figure}
\epsfig{file=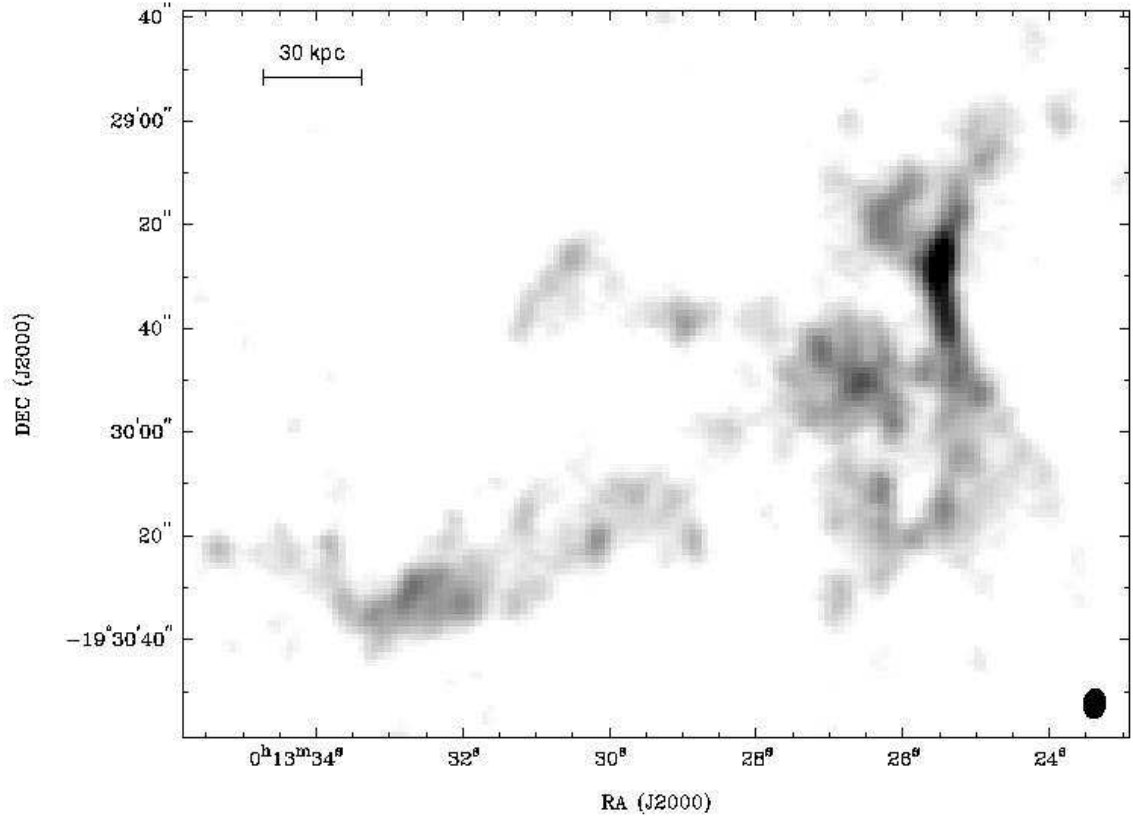,width=11cm,angle=-90}
\caption{
VLA naturally-weighted image of the relic in A13 at
1.425\,GHz. The FWHM restoring beam in the lower right corner is
5.9$''\times 4.4''$ in PA\,=\,$-$8\deg.  The surface brightness
varies from 40\,$\mu$Jy\,beam$^{-1}$ to 240\,$\mu$Jy\,beam$^{-1}$
(1.4\,K to 8.1\,K) and the r.m.s.\ noise level outside the source is
16.7\,$\mu$Jy\,beam$^{-1}$ (0.59\,K). The horizontal bar indicates the
linear scale, using the redshift from Table~1.
\label{fig1}}

\end{figure}

\begin{figure}
\epsfig{file=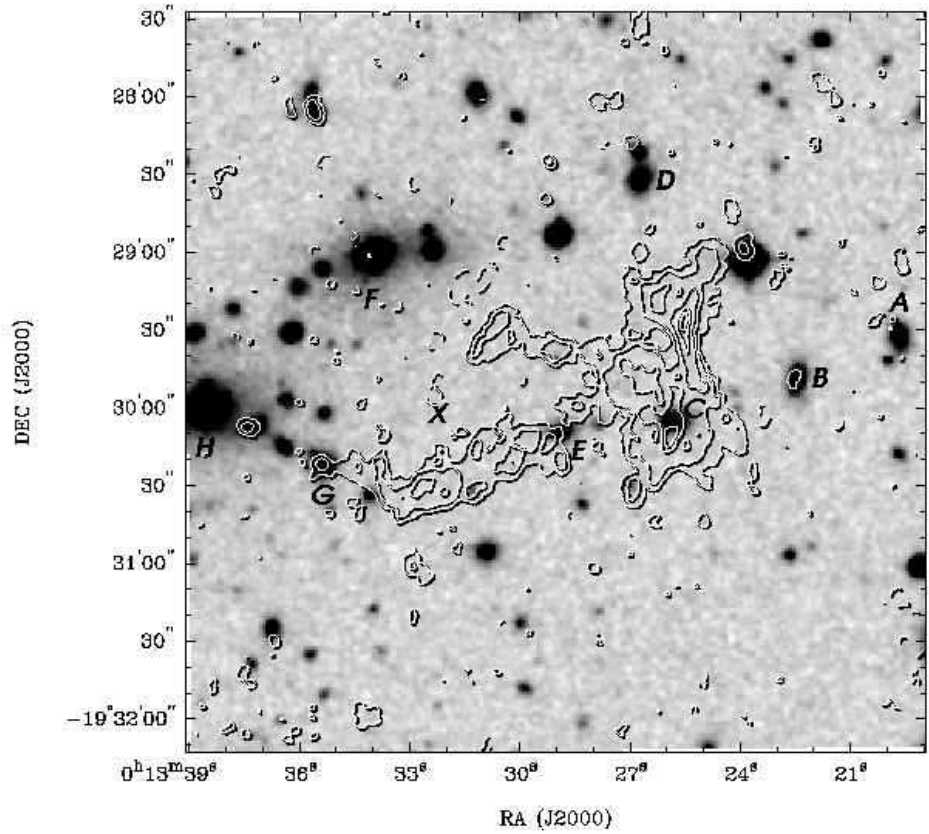,width=11cm,angle=-90}
\caption{
Radio contours of the relic in A13, corresponding to the grey-scale image 
in Fig.~\ref{fig1}, are overlaid on a red DSS-2 image of the area. 
The contour levels are ($-$43 (dashed), 43, 86, 173, 303, 
389)\,$\mu$Jy\,beam$^{-1}$ or ($-1.47, 1.47, 2.9, 5.8, 10.2, 13.0$)\,K.
The labelled galaxies are discussed in \S\ref{optids}. \label{fig2}}
\end{figure}

\begin{figure}
\epsfig{file=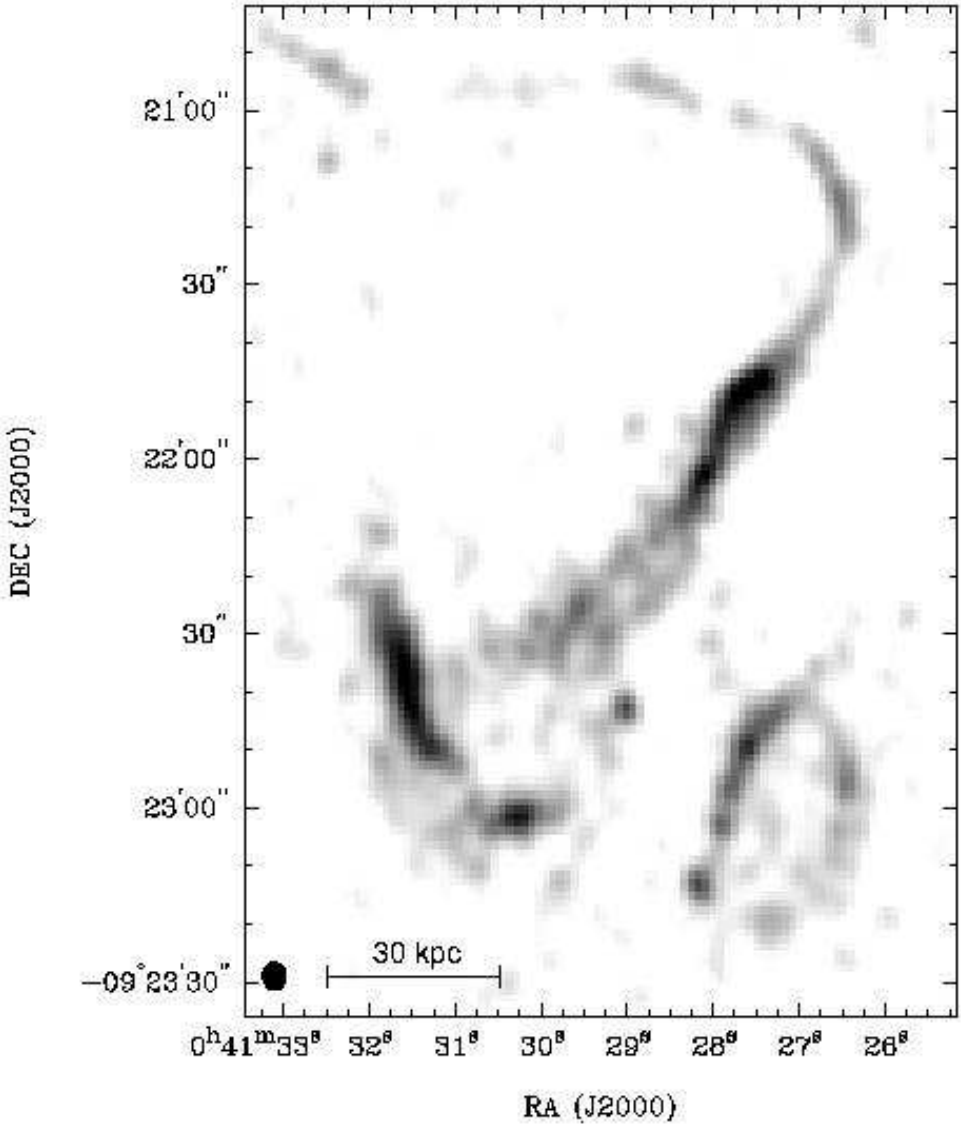,width=15cm,angle=-90}
\caption{ VLA uniformly-weighted image of the relic
in A85 at 1.425\,GHz.  The FWHM restoring beam in the lower left
corner is 5.0$''\times 4.2''$ in PA\,=\,5\deg.  The surface brightness
varies from 30\,$\mu$Jy\,beam$^{-1}$ to 300\,$\mu$Jy\,beam$^{-1}$
(1.3\,K to 12.5\,K), and the r.m.s.\ noise level outside the source is
17.4\,$\mu$Jy\,beam$^{-1}$ (0.75\,K).  The horizontal bar indicates the linear
scale, using the redshift from Table~1.
\label{fig3}}
\end{figure}

\begin{figure}
\epsfig{file=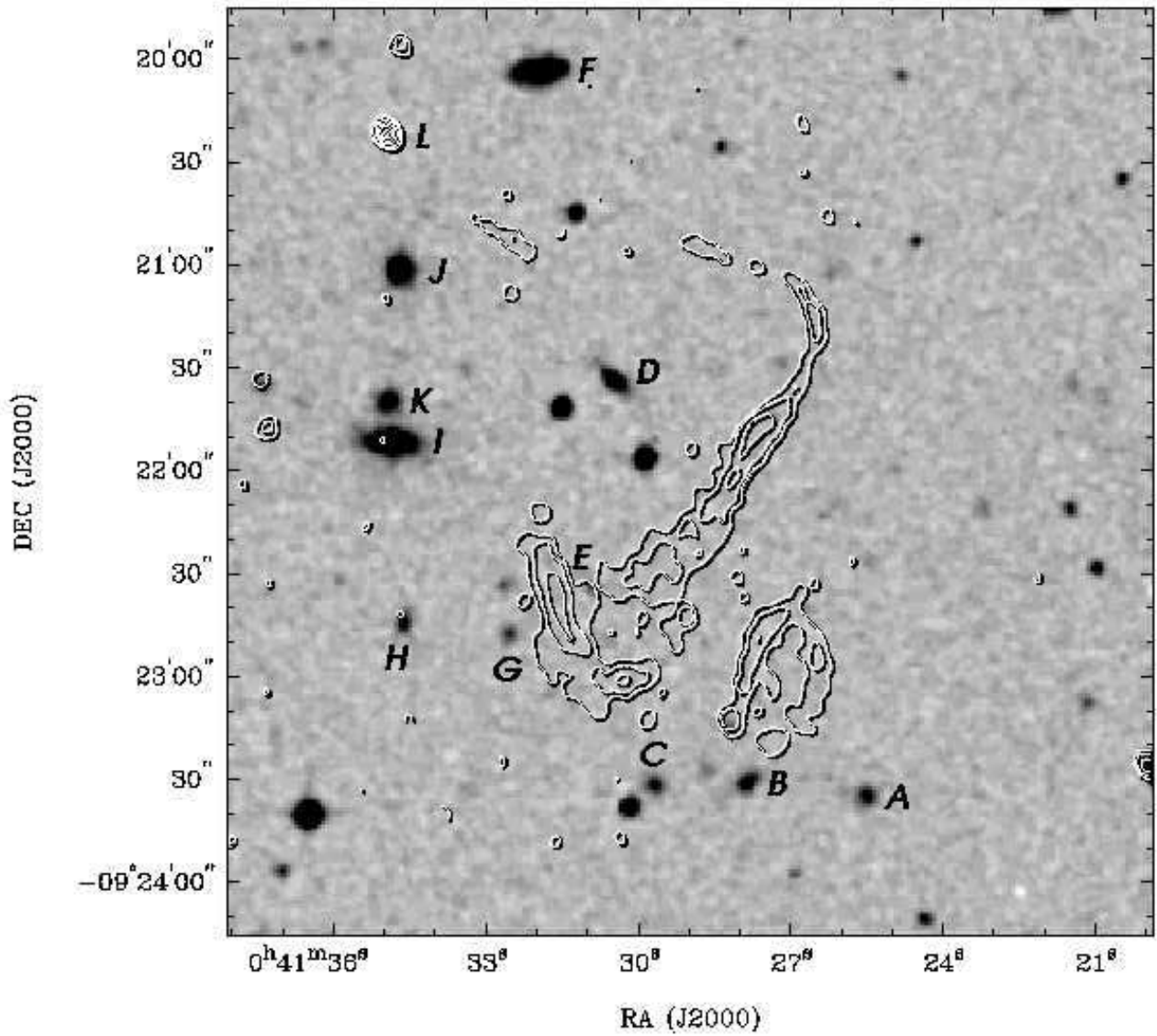,width=15cm,angle=-90}
\caption{
Radio contours of the relic source in A85, corresponding to the 
image in Fig.~\ref{fig3},
are overlaid on a red DSS-2 image of the area. 
The contour levels are ($-$43 (dashed), 43, 85, 171, 239, 
307)\,$\mu$Jy\,beam$^{-1}$ or $(-1.8, 1.8, 3.6, 7.1, 9.9, 12.7)$\,K. 
The labelled galaxies are discussed in \S\ref{optids}.
\label{fig4}}
\end{figure}

\begin{figure}
\epsfig{file=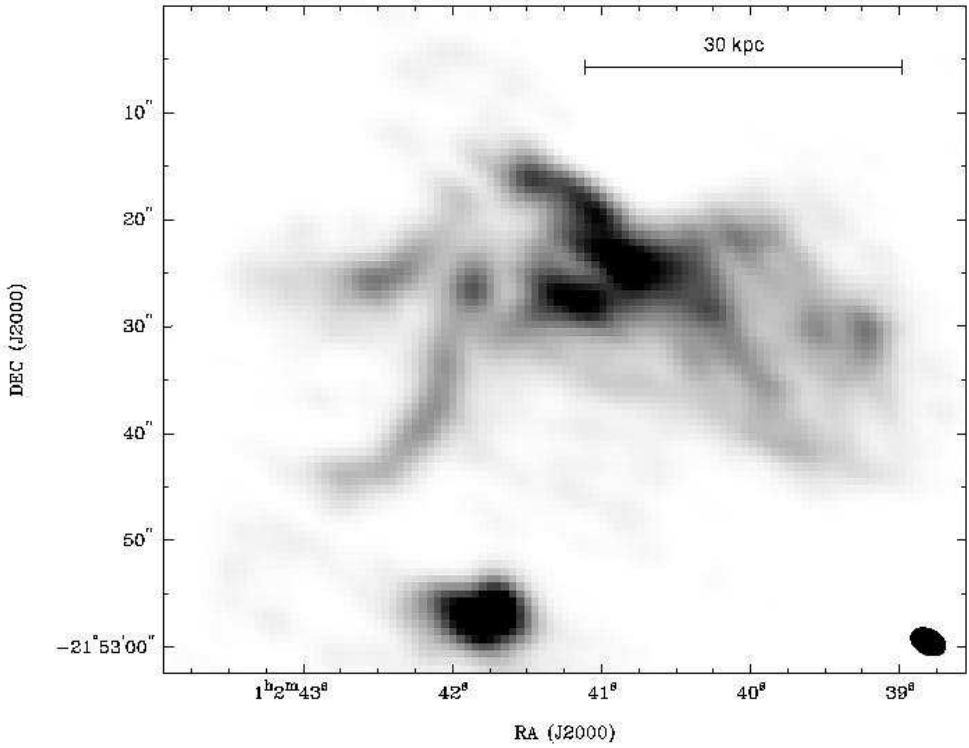,width=11cm,angle=-90}
\caption{
VLA uniformly-weighted image of the relic in A133 at
1.425\,GHz.  The FWHM restoring beam in the lower right corner is
3.5$''\times 2.4''$ in PA\,=\,64\deg.  The compact image at the bottom
is the source identified with the cD galaxy at the cluster
centroid. The surface brightness varies from 25\,$\mu$Jy\,beam$^{-1}$
to 3000\,$\mu$Jy\,beam$^{-1}$ (2.6\,K to 310\,K) and the r.m.s.\ noise
level outside the source is 21.1\,$\mu$Jy\,beam$^{-1}$ (2.21\,K).  The
horizontal bar indicates the linear scale, using the redshift from Table~1. 
\label{fig5}}
\end{figure}

\begin{figure}
\epsfig{file=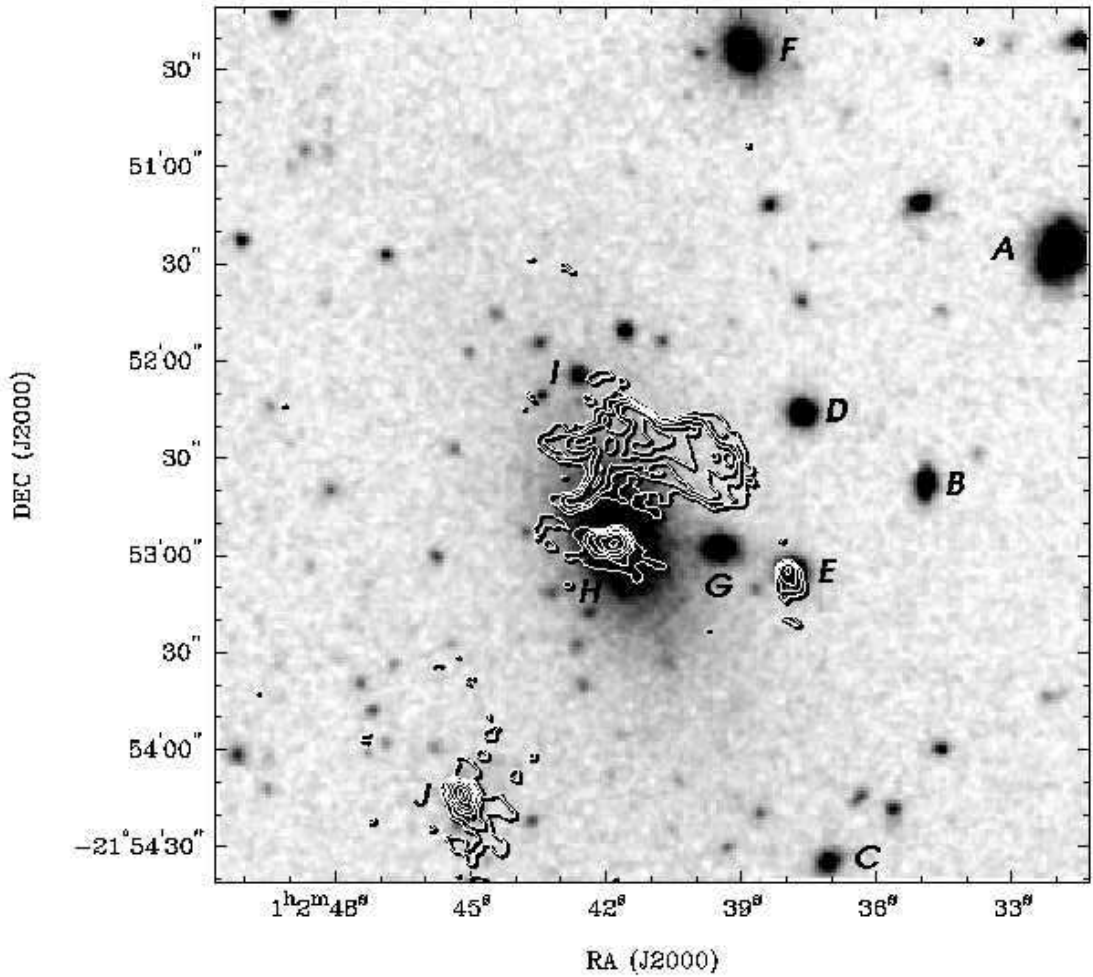,width=13cm,angle=-90}
\caption{
Radio contours of the relic source in A133, corresponding to the image
in Fig.~\ref{fig5}, are overlaid on a red DSS-2 image of the area. The
contour levels are ($-$95 (dashed), 95, 237, 475, 950, 2137, 3325,
4275)\,$\mu$Jy\,beam$^{-1}$ or $(-9.9, 9.9, 24.5, 49.1, 98.2, 221, 344,
442)$\,K.  The labelled galaxies are discussed in \S\ref{optids}.
\label{fig6}}
\end{figure}

Figs.~\ref{fig1}, \ref{fig3}, and \ref{fig5} show 
the relics in A13, A85 and A133, respectively.  
Figs.~\ref{fig2}, \ref{fig4}, and \ref{fig6} overlay the radio images
on red images from the Digitized Sky Survey (DSS-2).
The corresponding images of A4038 are shown by \cite{sr98}. 
The relics show a remarkable variety of fine structure that takes the
form of arcs, filaments and loops of enhanced surface brightness. Most
of the arcs and filaments are barely resolved (5\,kpc) in their
transverse directions, but can have projected lengths of up to
100\,kpc. The largest projected linear size of the relics (assuming
that they lie within the clusters) varies from 55\,kpc (in A133 and
A4038) to 260\,kpc (in A13). There is no consistent morphology among
the relics; in A133 (Fig.~5), four arc-like features appear to radiate
from the high-brightness centre of the relic, and in A13 and A85 the
structures have various orientations. The arcs and loop in A85, for
example, undergo large changes in direction, the loop in the
south-west structure showing a reversal of $\sim$\,180\deg. The relic
in A4038 (Slee \& Roy 1998) has a morphology that is most similar to
that of A85.

The A85 relic has been imaged by Giovannini \& Feretti (2000) at
333\,MHz with 20$''$ resolution. Their image shows low-brightness
extensions to the north-east (NE) and south-east (SE) of our 1.425 GHz
image, although a conspicuous void is centred south of galaxy~$D$ in
our Fig.~4. Galaxy $J$ lies within the tip of the NE extension.  Our
high-resolution image shows some contribution from the NE extension,
and also from the SE extension when we lower the grey-scale image to the
2 $\sigma$ level rather than the 3 $\sigma$ level of Fig.~3.

We compute from Giovannini and Feretti's average surface brightness of
about 20~mJy\,beam$^{-1}$ at 333\,MHz that we would not have detected
these extensions at the 3\,$\sigma$ level if the spectral index of
this diffuse radiation were $\ge$\,2.1.  Since they measure a spectral
index over the whole relic of $\alpha$\,=\,2.5 to 3.0, which is steeper
than this limit, we do not expect to detect the extensions seen in
their image because it will be below our brightness sensitivity.
Thus, it is not necessary to invoke a lack of low spatial frequencies
in our high-resolution image to explain the apparent absence of the
diffuse emission in Fig.~3.

The extensions are also seen at poorer resolution in the 1.4 GHz image
from the NRAO-VLA Sky Survey (NVSS, Condon et al.\ 1998),
which has better brightness sensitivity than our high-resolution
image. This diffuse emission must still be contributing to our measured
flux density in Table~2, which shows that there is no significant deficit
in our value with respect to NVSS.

Rizza et al.\ (2000) suggest that their less-sensitive,
lower-resolution VLA image of the relic in A133 shows a narrow
bridge of emission between the cD and the relic with brightness
$\sim$\,1\,mJy\,beam$^{-1}$ at 1.4\,GHz.  They interpret this as a
perturbed jet propagating through a cooling flow, giving rise to an
amorphous radio lobe. Our image (Fig.~5) does not support
the presence of this jet. The average 1.425 GHz brightness in the gap
between the south-east filament (which itself is curving away from the
cD) and the cD is only 26\,$\mu$Jy\,beam$^{-1}$, not far above the rms
noise level in regions well away from radio sources. When we convolved
our image with Rizza et al's beam, there was still no convincing
connection of the form that is shown in their image.

The referee suggests that the relic in A133 might actually be the lobe
of a normal radio galaxy whose jet is approaching us at a small angle
to the line of sight, and that the counterlobe lies in the south-east
part of the NVSS image.  However, the suggested counterlobe in the SE
of the NVSS image is predominantly due to the source identified with
galaxy $J$ in Fig. 6.  The relic and the source coincident with the cD
could be the approaching and counter lobes of a radio galaxy (with the
cD as the host), provided one is prepared to accept that both lobes
possess unusually steep spectra (spectral indices of 2.1 in Table~2),
and that the lobes differ by a factor of $\simeq 6$ in flux density
and linear size. Since other cluster radio galaxies do not possess
such extreme differences in lobe parameters (Slee et al. 1998, Table
2), we propose that the relic is not the lobe of a conventional radio
galaxy that is associated with the cD.  The referee further suggests
then that the relic could be a wide-angle tail radio galaxy seen close
to side-on and so is seen with the tails projected onto each other,
although again the observed spectral index is unusually steep for a
conventional radio galaxy.  One might be able to test this proposal
observationally since the relatively long lines of sight through the
radio-emitting plasma should produce a large Faraday depolarization.

\subsection{The Flux Densities and Radio Spectra}  \label{fluxspec}

\subsubsection{The High-Resolution Images}  \label{highres}

We measured the integrated flux density within a rectangular box
surrounding each relic and, to estimate the uncertainties due to
thermal noise and residual sidelobes of incompletely cleaned field
sources, we executed the process 11 times, varying the box sizes
and locations around the relic centroids.  Systematic errors were
estimated to be 3\,\% for the flux-density scale, 1\,\% for the flux-density
bootstrapping and 3\,\% for the deconvolution, estimated by repeated
deconvolution with various values of the deconvolution control parameters.
When added in quadrature, the total uncertainty on our flux-density
measurements was 4.5\,\% to 5.0\,\%, and the resulting flux densities
are given in Table~2.

To estimate the integrated spectral index, we measured the integrated
flux density over the relic 11 times at 1.385\,GHz and 1.465\,GHz
within a rectangular box that was varied in area by a factor of four
and with box centers that were offset in the E-W and N-S directions
from the relic centroid, but always included the whole of the relic.
From each pair of measurements we derived the spectral index.  Thus we
made 11 experimental observations of a constant value, and the
measurements differ slightly from each other due to differing
contributions of thermal noise, and sidelobes, depending on the box
size.  Systematic errors are virtually identical at both frequencies.
From the 11 measurements we determined the sample mean.  The precision
of the sample mean was estimated by the standard error of the mean.
The integrated spectral indices at the central frequency of 1.425\,GHz
were in the range 2.1 to 4.4.

We estimated the potential missing flux density due to resolution
effects by measuring flux densities from the images in the NVSS.
The survey has a beamwidth of
$45''$ and so suffers less from missing flux density but suffers more from
confusion with nearby unrelated sources.  We measured flux densities
from the NVSS images in the same manner as explained above, but then
subtracted the contributions of unrelated sources that are clearly
seen and measurable in our higher-resolution images.
These 1.40 GHz flux densities from the
NVSS are listed in Table~3 and are discussed further in
\S\ref{broadspec}.  The flux density missed from 
each relic by our present high-resolution observations was computed 
by subtracting the flux-density measurement at high resolution (Table~2;
corrected for the frequency difference) from the flux-density measurement
from the NVSS,
and the largest difference amounted to 17\,\% of the corrected NVSS
flux density.  The flux-density differences are given in Table~2.  The
uncertainties in these values are large because they are small
differences between large quantities.  Our flux-density estimates
are not significantly lower than those of the NVSS.
 
We attempted to measure the variation in spectral index over each
relic, but found when we subdivided the relic, the signal-to-noise
ratio dropped to such an extent that, with a frequency base of only 
80~MHz, the uncertainty on the spectral indices was too large for the
spectral index measurements to be useful.

\subsubsection{The Integrated Spectra} \label{broadspec}                

Additional flux-density measurements at other frequencies from the
literature were used to plot the broad-band spectra of the relics in
Figs.~8 and 9, using the data in Table~3. The details of these
measurements are given in the following notes, which are numbered in
column~10 of Table~3.

(1) The 16.7\,MHz measurements were made by Braude et al.\ (1981) with
a FWHM beam of 40$'\times$80$'$.  The excess flux density included in
their value (column~3) from the many other sources within their beam
was estimated by summing the flux densities from the 1.4\,GHz sources
in NVSS that fell within this beam and extrapolating that value to
16.7\,MHz with an assumed average spectral index of 0.80 $\pm$ 0.1;
the error in the excess flux density reflects the range of assumed
spectral indices. Braude et al.\ made an absolute calibration of their
flux-density scale, which is close to the extrapolated scale of
``KPW'' (Kellermann, Pauliny-Toth \& Williams 1969).  The authors give
a flux-density error that includes the effects of noise and confusion,
although they could not have included the contributions of the many
sources within their beam, as we have outlined above.

(2) The 29.9\,MHz measurements of Finlay \& Jones (1973) were made with a
FWHM beam of  50$'\times$50$'$ at the zenith. The corrections for the excess 
flux densities of the
numerous sources within their beam (columns 3, 5, 7) were made as explained
above. The authors made an absolute calibration of their flux-density scale, 
which they found was $\sim$\,5\,\% above the extrapolated KPW scale.

(3) The 80\,MHz measurements of Slee (1995) were made with a FWHM beam
of 3.75$'$ at the zenith. The only correction for confusion necessary
was for the source A133\_8 (Slee et al.\ 1994, present paper), which
has a well-determined spectrum. The flux-density scale was that of KPW and the
random errors (Slee \& Higgins 1973) depend on the flux density and
number of measurements contributing to the average. The total error
varied between 12\,\% and 20\,\%, including a contribution of 10\,\% for 
uncertainty in the flux-density scale.

(4) The 160\,MHz measurements of Slee (1995) were made with a FWHM beam of
1.85$'$ at the zenith. No corrections for confusion were necessary. The
flux-density scale was that of KPW and the errors are discussed in Slee (1977).

(5) The 408\,MHz measurements of Reynolds (1986) and Slee \& Reynolds
(1984) were made with a FWHM beam of 2.6$'\times$2.9$'$ at the zenith.
Small corrections for confusion were needed for A133\_8, and
A4038\_10, A4038\_11 (Slee et al.\ 1994) using a flux density
extrapolated from 1.5\,GHz.  An absolute calibration of the 408\,MHz
flux-density scale was made by Wyllie (1969), and Reynolds (1986)
suggests that errors of $\sim$\,15\,\% are appropriate for these flux
densities.

(6) The 843\,MHz measurements of Reynolds (1986) and Slee and Reynolds
(1984) were made with a FWHM beam of $40'' \times 40''$\,cosec($\delta$).
Corrections for confusion were necessary only for A4038\_10 and
A4038\_11 (Slee et al.\ 1994) using the extrapolated flux densities
from 1.5\,GHz.  Reynolds (1986) suggests that errors of $\sim$\,7\,\%
are appropriate.

(7) The 1.4\,GHz flux densities from NVSS (Table 3) were used rather
than our measured values from Table~2, to include any flux density
that our high-resolution measurements may have failed to detect.  The
NVSS FWHM beam of 45$''$ includes the sources A133\_8, A4038\_10,
A4038\_11; we have corrected the NVSS total flux densities by
subtracting the VLA flux densities given in Table 2 for those
sources. The NVSS flux-density scale is that of Baars et al.\ (1977)
and has an error of 3\,\% to 4\,\%, and the combined error is unlikely
to exceed 5\,\% to 10\,\%, depending on the flux density.

(8) The 2.7\,GHz measurements of Andernach et al.\ (1986) and Reuter
\& Andernach (1990) were made with a FWHM pencil beam of 4.4$'$. A
substantial correction was made for A133\_8 by extrapolating from our
present 1.425\,GHz flux using the known spectral index from Slee et
al.\ (1994). The flux-density scale is that of Baars et al.\ (1977)
and the total error quoted by the authors for the relic in A133 was
30\,\%.

(9) The 4.9\,GHz measurement of the relic in A133 from Slee et al.\
(1994) was made with the VLA with a FWHM beam of
21$''\times$14$''$. No corrections for confusion were needed. The
flux-density scale was that of Baars et al.\ (1977) and the total
error was 7.5\,\%. The upper limit for A4038 was based on multiplying
the 3 $\sigma$ brightness limit by the number of beam areas subtended
by the source.

We believe that the flux densities in Table 3 are free of errors
associated with the inclusion of field sources within the beam of
the low-resolution telescopes, and that the quoted errors include 
the uncertainties in the flux scales adopted by various observers.

The broad-band spectra in Figs.~8 and 9 show a characteristic curvature
with pronounced flattening at the low-frequency end; at frequencies below 
$\sim$\,100\,MHz the slopes of the spectra are flatter than $\alpha$ = +1, 
whilst near 1.425\,GHz the spectra are much steeper with 
spectral indices given in Table~2. The significance of the 
high degree of curvature seen in these spectra will be explored 
using spectral-ageing models developed in \S\ref{specanal}.

As a consistency check, we compared the spectral index at 1.425\,GHz
from the broad-band spectra to the spectral index at 1.425\,GHz from
the high-resolution images, using model fits to the broad-band spectra 
from \S\ref{specanal} to interpolate between the data points of the
broad-band spectrum taking into account the spectral curvature over a
broad frequency range.  We found serious discrepancy only for A13.
Details of the comparison are in \S\ref{sumanal}.

\subsection{Polarization} \label{polari}   

When estimating the average polarization fraction, we subdivided each
relic into between ten and 16 approximately equal regions, each
containing about 20 beam areas.  The polarized flux density and total
intensity were measured for each region, resulting in ten to 16 values 
for the polarized fraction.  The average polarization fraction over
each relic was determined using the sample mean.  The
polarization fraction varies systematically within the relics by an
amount that is large compared to the random noise in the image and so,
in contrast to the case of the spectral index measurement, the sample
standard deviation quantifies an interesting property of the relics,
that is, how much the polarization fraction differs from the mean from
region to region.  We quantify this variation by using the standard
deviation.  The average polarization percentages with their standard
deviations are listed in Table~2. The position angle of the electric
vector at the source is not known for any of these relics because we
lack the higher-frequency image with similar angular resolution with
which to determine the rotation measure.

The high value of the standard deviation of the polarization fraction
in A85 reflects a significant variation in different parts of the relic.
Figure~\ref{fig10} shows that the polarization varies between
$\sim$\,35\,\% in the north-west area, $\sim$\,25\,\% in the
south-west loop and 5\,\% to 10\,\% in the south-eastern
arc. There is no clear correlation between polarization and total intensity.
Unfortunately, the lack of comparable accuracy in the measurement
of spectral index prevents our searching for a relationship between
polarized fraction and spectral index. 

The polarization varies significantly among these three relics. A85 is
highly polarized in some areas at 1.425\,GHz, implying that the fields
are not highly tangled and that Faraday depolarization is low.
Conversely, A133 and A4038 have low fractional polarization.  Whether this is
due to tangled fields and/or Faraday depolarization can be determined
only when a higher-frequency image of similar angular resolution
becomes available.

\subsection{Optical Identifications} \label{optids}

The identification of galaxies with which these relics may be
associated is an uncertain pursuit, but one can make some progress.
First, if the relics are the remnants of conventional radio galaxies,
then the former host should be a bright elliptical with
$M_{\mathrm{R}}\le -21.0$ (cf.\ Ledlow \& Owen 1995; Slee, Roy, \&
Andernach 1998).  Second, the galaxy should be close enough to the
relic that it could have moved to its present distance 
in the lifetime of the relic.  The brightest cluster member (BCG) is
naturally the prime suspect, provided that it satisfies
the distance criterion.
Note, however, that these travel-time arguments are uncertain because of
the use of projected spacings and velocities, which vary respectively as
the cosine and sine of the angle with respect to the plane of the sky.
The distance criterion can also be affected strongly by intracluster winds, 
which are known to shape the tails of many tailed sources in clusters
and which might exceed 1000\,km\,s$^{-1}$ (Burns 1998).  Such a wind 
could carry
a relic away from the host faster or slower than expected from the
peculiar motion of the host galaxy, depending on the vector addition 
of wind and host velocities.  We have, unfortunately, no knowledge of 
wind speeds or directions in any of these four clusters, and so
the travel times we derive are probably rather uncertain.
The identification should also allow for displacements between the
host galaxy and radio centroid that are commonly seen in radio galaxies
in clusters.  For example, the median differences in projected positions
between the radio centroid and optical host in the cluster sample of
Slee et al. (1994, 1996) yields medians of 5'' for 25 doubles, 13'' for 
seven wide-angle-tail sources and 13'' for 13 narrow-angle-tail sources,
at an angular resolution of 14''.

A13 - The galaxies identified in Fig. 2 have redshifts compatible with
membership of A13 and satisfy the luminosity criterion for a former
host galaxy, except galaxy $E$ for which we have no redshift
measurement.  Travel times for these galaxies from the centroid of the
radio relic (10$''$ NW of galaxy E) range from $0.58 \times 10^8$\,yr
(galaxy $C$) to $6.6 \times 10^8$\,yr (galaxy $B$), assuming that the 
velocity vector is at $45^{\circ}$ to the plane of the sky.  Only galaxy $C$
has a travel time that is within a factor of two of the relic's age of
$t_{\rm RE} = 0.33 \times 10^8$\,yr from the KGKP model of
\S\ref{specanal}; the remaining galaxies have travel times $\ge 1.1
\times 10^8$\,yr, which are better matched by the relic age derived using the 
MJP model ($t_{\rm RE} \ge 2.44 \times 10^8$ yr).  However, the differences 
between relic
ages from the KGKP and MJP models make it difficult to use the travel
times alone to select one of the bright galaxies as a former host.

One could argue that galaxy $C$ is the active host to a source reminiscent of
wide-angle-tail radio galaxy with tails stretching to the north and
north-east.  If so, then the linear structure stretching to the
east-south-east could equally well be interpreted as another tailed radio
galaxy with active host either galaxy $E$ or $G$ (though it is not certain that
$E$ is in A13).  However, the extreme spectral indices of both radio
structures ($\simeq 4.4$) make it very unlikely that these are active radio
galaxies. 

The BCGs ($F$ and $H$) are the preferred hosts because
optically luminous galaxies are the most likely hosts to radio
galaxies (eg Slee et al. 1998, Fig. 18), and the travel times ($1.7
\times 10^8$\,yr and $1.2 \times 10^8$\,yr respectively) lie within
the range of relic ages estimated by the KGKP model ($t_{\rm RE} = 0.33
\times 10^8$\,yr) and the MJP model ($t_{\rm RE} \ge 2.44 \times
10^8$\,yr).

A85 -- The BCG (a cD) lies 395\,kpc to the north-east of the relic's
centroid and is well outside the optical overlay in Fig.~4. Its low
peculiar velocity ($\le$\,100\,km\,s$^{-1}$) results in a
travel time from the relic of $\ge$\,4$\times$10$^9$\,yr (assuming that 
the velocity vector is at $45^{\circ}$ to the plane of the sky), which would
make the relic unreasonably old, unless the relic is due to recompression
(En{\ss}lin \& Gopal-Krishna 2001).

The brighter galaxies within 2.5$'$ (160 kpc) of the relic's centroid
are labelled in Fig.~4. All have measured redshifts that place them in
the cluster, except galaxies $I$ and $F$ which are foreground 
and have recession velocities that differ from the cluster velocity
by two and a half and three times the cluster velocity dispersion, and
galaxies $C$, $E$, and $G$ for which no redshifts have been measured.
Of the galaxies known to be in the cluster, only $J$ barely fulfils
the luminosity criterion for a radio galaxy, having $M_{\mathrm{R}} =
-21.0 \pm 0.2$.  It lies near the north-eastern tip of a
low-brightness extension of the relic, which is more clearly detected
in the 333\,MHz image of Giovaninni \& Feretti (2000) than in our 
high-resolution images.  (However, note that
the 333\,MHz image confuses the compact radio source marked $L$ in Fig.~4 with
the relic.)  If $J$ is the former host, its projected distance from
the relic centroid (128\,kpc) and its peculiar radial velocity
(676\,km\,s$^{-1}$) yield a travel time from the relic of
1.8$\times$10$^8$\,yr, which is within a factor of two of the spectral age
($1.0\times10^8$\,yr) derived for the relic in \S\ref{specanal}.

We conclude that if the relic was a radio galaxy and a member of A85,
then its former host is most likely to be galaxy~$J$, which is still
embedded in a diffuse relativistic plasma visible at 333\,MHz.\\

A133 -- the brighter galaxies within 2.5$'$ of the relic's centroid
are labelled in Fig 6. The galaxies $B, C, E, F, G,
H,$ and $I$ possess measured redshifts and those are consistent with
cluster membership. Galaxy $J$ is identified with a radio source
(A133\_8 in Slee et al.\ 1994) and has a redshift of 0.2930, which
places the galaxy well behind the
cluster. This source is of interest because it is not properly
resolved (if at all) from the relic and the cD in the lower-resolution
observations and its contribution has been subtracted in Table 3.

Galaxy $H$ (the cD ) would normally be favoured as the former host, but the
travel time to its present position of $1.5 \times 10^8$ yr (assuming that 
the velocity vector is at $45^{\circ}$ to the plane of the sky) is three 
times the relic's age of $t_{\rm RE} = 4.9 \times 10^7$ yr (\S\ref{specanal}). 
The remaining five
ellipticals in Fig. 6 that have measured cluster redshifts and are bright
enough to qualify as host galaxies ($B, C, E, F,$ and $G$) possess travel times
that range from $4.2 \times 10^7$ yr (galaxy $G$) to $1.6 \times 10^9$ yr 
(galaxy $E$). The travel time of galaxy $G$ most closely matches the 
age of the relic.

In summary, the cD is unlikely to be an active host to a radio galaxy, of
which the relic forms one lobe, but it could be the former host, provided
one is prepared to accept that the calculated travel time and/or relic age
have rather large uncertainties.

A4038 -- The brighter galaxies within 4$'\times$4$'$ of the relic's
centroid are labelled in Fig. 7.  The galaxies $A, B, D, E, F, G, H,
I,$ and $J$ have measured redshifts that are consistent with cluster
membership.  The image marked $C$ is stellar. The galaxies $A, D, E,
G, H,$ and $J$ satisfy the luminosity criterion for a radio galaxy
host.  The cD (galaxy~$H$) is the most likely candidate due to its
large luminosity.  It is currently host to a radio galaxy, and so if
it produced the relic, it is now undergoing a second stage of
activity.  If the velocity vector lies at $45^{\circ}$ to the plane of
the sky, it would have reached its present position from the relic
centroid in $3.1 \times 10^8$\,yr.  
Travel times for the remaining five cluster galaxies range from 
$4.5 \times 10^7$\,yr to $5.5 \times 10^8$\,yr. Galaxies A,E,G and J have 
travel times within a factor
of two of the relic ages $t_{\rm RE}$ deduced from the spectral modeling in
\S\ref{specanal}, but galaxies D and H (cD) exceed the model ages by 
factors of at least three to five.

\section{Relics and X-ray Emission}  \label{relix}

Since the pressure of the hot intra-cluster gas is likely to play a
significant role in the evolution of cluster relics, it is important to
compare the morphologies of radio and X-ray images to search for enhanced
X-ray emission from the regions of the relics.

A connection between magnetic fields, cosmic-ray electrons and the
cluster X-ray emission recently received further observational support 
when both Rephaeli, Gruber \& Blanco (1999) and 
Fusco-Femiano et al.\ (1999) independently reported the 
discovery of an additional hard X-ray component in the Coma
cluster emission with the Rossi X-ray Timing Explorer and BeppoSAX, 
respectively.  Although there is
some confusion from the presence of the Seyfert 1 galaxy (X Comae) in the
wide field of view of their instruments, both groups find evidence for
non-thermal inverse-Compton X-ray emission as predicted for clusters with
extended radio sources (Rephaeli 1979).  More recently, 
En{\ss}lin et al. (1999), Dogiel (2000), and Liang (2001) find that the
high-energy excess is not likely to be inverse-Compton emission but
rather is bremsstrahlung from a non-thermal tail of the electron energy 
distribution.  However, Petrosian (2001) disagrees and concludes that the 
background thermal electrons cannot be the source of the non-thermal
electrons, except for a short period of less than $10^8$ yr.

Pointed X-ray observations are available in the ROSAT archive for A85,
A133 and A4038, but for A13 we have only the XBACs data of Ebeling et al.
(1996), based on the ROSAT All-Sky Survey (RASS). Our X-ray analysis was
performed using the EXtended Scientific Analysis System EXSAS (Zimmermann
et al.\ 1994). Several observations were merged resulting in 15.9 ks PSPC
and 31.7\,ks HRI exposures for A85 and A133, respectively. In the case of
A4038 a single 3.3\,ks PSPC data set was available. Binning the PSPC data to
a pixel size of 5$''$, we produced images of the central part of the detector
window out to 20$'$ from the optical axis. These were exposure corrected and
an energy-dependent Gaussian filter was applied to compute X-ray color
images in the broad (0.1\,keV to 2.4\,keV), soft (0.1\,keV to 0.4\,keV) and 
hard (0.5\,keV to 2.0\,keV) ROSAT energy bands. 
Details of the image production are given in Ehle et al. (1998).
The PSPC hard-band images of A85 and A4038 are
shown as contour plots overlaid on grey-scale radio images of the relic
sources in Figs.~11a and 13, respectively. 

To minimize the HRI instrument
background on A133 (due to UV emission and cosmic rays) without degrading
the X-ray sensitivity, we selected raw amplitudes from only channels 2 to 8
for our analysis (see David et al.\ 1993). The HRI image was constructed
with a pixel size of 1$''$ and smoothed with a Gaussian filter of 12$''$ FWHM.
The HRI contour map of A133 is overlaid on a grey-scale radio image of the
relic source in Fig.~12a.

To improve contrast in regions of excess X-ray
emission in the region of the relic, we attempted to subtract
the cluster thermal emission component from A85, A133 and A4038.
For A85, we achieved the lowest residuals by subtracting a 
simple circularly-symmetric Gaussian model consisting of three components 
that had a common centre point, located at the peak of the cluster X-ray 
emission, and that had differing widths and amplitudes.
(The FWHM of the components were 110'', 310'', and 750'', and the relative 
peak heights were 1.0, 0.168, 0.076.)  For A133 
we subtracted a mask created by azimuthally-averaging the
image of the cluster X-ray emission.  The centre of rotation was the
peak of the cluster emission, and the lower half-image only was used
since we found that including the relic emission in the mask resulted
in over-subtraction of the region south of the X-ray peak.  
For A4038 we tried subtracting various models centered on the X-ray peak,
which itself was offset from the cD in the direction of the relic.
No convincing feature in the residual image was associated with the relic.
The resulting residual images for A85 and A133 are shown in Figs.~11b and 12b.

The plots in Figs.~11, 12 and 13 show a slightly enhanced diffuse
X-ray emission in the region of the radio source (A85 and A133) or a
shift of the X-ray emission centroid towards the relic (A4038). The
excess X-ray emission seen in Figs. 11b and 12b is most pronounced 
in the hard energy band, as can be
expected for inverse-Compton emission, which has a flatter spectrum than 
the cluster's thermal emission, more in keeping with the radio spectral
indices present in the low-frequency parts of Figs.~8 and 9. We note
that in the case of A85, enhanced X-ray emission in the relic region
was also reported by Bagchi, Pislar \& Lima Neto (1998). However,
there appears to be a difference between the spectral composition of
excess X-ray emission seen in the Coma halo and the present
relics. Our relics are associated with excess emission in the energy
range 0.5\,keV to 2\,keV, while the excess in Coma is detected only
above 18\,keV. This could be explained most easily by a difference in
the very-low-frequency spectral indices of halos and relics; relics
could have a higher density of relativistic electrons radiating at the
lower frequencies that most efficiently scatter CMB photons to our
lower energy X-ray spectral band.

The excess X-ray flux from the region of the radio relic, measured
from the residual X-ray images is 
$2.05 \times 10^{-12}$\,erg\,cm$^{-2}$\,s$^{-1}$ for A85 and
$8.97 \times 10^{-13}$\,erg\,cm$^{-2}$\,s$^{-1}$ for A133, estimated
for the 0.5 keV to 2.4 keV band and corrected for Galactic H\,I absorption.
Integrating the radio flux from Table 3 between 10 MHz and 100 MHz yields
integrated fluxes of $5.10 \times 10^{-17}$\,W\,m$^{-2}$ for A85 and 
$3.61 \times 10^{-17}$\,W\,m$^{-2}$ for A133.  Assuming that the X-ray
flux is from inverse-Compton scattering, the magnetic field inferred
following Bagchi et al. (1998) is 0.47\,$\mu$G for A85 and 0.60\,$\mu$G for
A133.  The value for A85 differs by a factor of two from the 
($0.95 \pm 0.10$)\,$\mu$G found by Bagchi et al.
At least a part of this difference is probably due to
differences in the details of calibrating and measuring the X-ray and 
radio fluxes.

Confirmation of the presence of inverse-Compton emission in radio
relics will only come from the present generation of X-ray telescopes
such as {\it XMM-Newton} and {\it Chandra}.  These have the angular
and spectral resolution to match those of the present radio images,
and are sensitive to higher energies than ROSAT, which should provide
better contrast between inverse-Compton emission and the cluster
thermal emission.

\section{Spectral Modelling}   \label{specanal}  

The shape of the radio spectrum from 10 MHz to 5 GHz results from the
competition between energy injection and losses due to adiabatic
expansion, synchrotron emission, and inverse-Compton scattering off
cosmic microwave background (CMB) photons, and a detailed understanding
of these processes can yield the age of the relic, the duration of
initial particle injection, and the magnetic field strength.

\subsection{The Komissarov \& Gubanov Models}  \label{komgub}

Well-developed models of relic sources have been published by Komissarov \&
Gubanov (1994); they include a discussion of the relics  in A85 and A133
that are presented in this paper. The authors explored the possibility that
these extremely steep-spectrum radio sources could be the remnants of
powerful radio sources whose nuclei have ceased to produce jets.
They assumed a finite duration for the continuous injection (CI) of
relativistic electrons, $t_{\rm CI}$, followed by a relic (RE) phase of 
duration $t_{\rm RE}$ during which the injection of relativistic electrons
is switched off.
During both the CI and RE phases, the electrons lose energy by
synchrotron emission and inverse Compton (IC) scattering off 
CMB photons.
They supposed that the magnetic field in the lobes is uniform and constant
and calculated the broad-band synchrotron spectrum by modifying
the classical Kardashev-Pacholczyk (KP) model (Kardashev 1962)
and the Jaffe-Perola (JP) model (Jaffe \& Perola 1973).
Both the KP and the JP models consider an isotropic injection of
electrons.
The KP model describes the situation in which the time-scale
for continuous isotropization of the electrons is much  longer than the
radiative lifetime, each electron maintaining its original pitch angle
with respect to the magnetic field. Since the synchrotron losses depend 
on the magnetic field strength, $B$, and on the pitch angle, $\theta$, as  
$(B~{\rm sin}\,\theta)^{2}$, the small pitch-angle electrons save their
energy and the distribution becomes anisotropic.

In contrast, the JP model allows for very efficient pitch angle scattering 
and the electron distribution stays isotropic.
If inverse-Compton losses are taken into account, the energy $\epsilon$ of
 a single electron evolves at a rate given by
\begin{equation}
\frac{d\epsilon}{dt}=-(b_{\rm syn}+b_{\rm IC}) \epsilon^{2} 
\end{equation}
where factors $b_{\rm syn}$ and $b_{\rm IC}$ describe synchrotron and inverse-Compton losses, respectively. 
 The electron's {\it synchrotron} losses for the KP model and the JP model, are given by
\begin{center}
$b_{\rm syn}({\rm KP})=c_{2} B^{2}(\sin^{2}\theta)$ \\[1.5ex]
\end{center}
 and
\begin{center}
$b_{\rm syn}({\rm JP})= c_{2} B^{2}\cdot (2/3)$ \\[1.5ex]
\end{center}
where $c_{2}=2.37 \cdot 10^{-3}$ erg$^{-1}$ s$^{-1}$ gauss$^{-2}$ is a 
constant (Pacholczyk 1970).
Because of the isotropy of the CMB radiation field, the inverse-Compton-loss
term is independent of pitch angle for both the JP and KP 
models  
\begin{center}
$b_{\rm IC}({\rm CMB})= c_{2} B_{\rm CMB}^{2}\cdot (2/3)$ \\[1.5ex]
\end{center}  
$B_{\rm CMB}=3.25\cdot(1+z)^{2}$ ${\rm \mu G}$ (where $z$ is the source redshift) is the magnetic field strength with energy density equal to that of the CMB.
Hence, adopting an effective IC magnetic field of $B_{\rm IC}=\sqrt{2/3}\cdot B_{\rm CMB}$, we combine synchrotron and inverse-Compton losses according to
 \begin{center}
$b_{\rm syn}({\rm KP})+b_{\rm IC}=c_{2} B^{2}(\sin^{2}\theta+B_{\rm IC}^2/B^{2})$ \\[1.5ex]
$b_{\rm syn}({\rm JP})+b_{\rm IC}=c_{2} B^{2}(2/3+B_{\rm IC}^2/B^{2})$. \\[1.5ex]
\end{center}

Here we refer to the anisotropic and to the isotropic 
Komissarov \& Gubanov  model as the KGKP and KGJP models, respectively.\\

The KGKP model is characterized by five free parameters:
\begin{itemize}
\item[1)] the injection spectral index of the CI phase, $\alpha_{\rm inj}$,
\item[2)] the total source age, $t=t_{\rm RE}$+$t_{\rm CI}$,
\item[3)] the relative duration of the relic phase with 
  respect to the CI phase, $t_{\rm RE}$/$t_{\rm CI}$,
\item[4)] the ratio between the source magnetic field strength and 
 the effective IC magnetic field, $B/B_{\rm IC}$,
\item[5)] the flux normalization.  
\end{itemize}

In the emission spectrum three break frequencies appear: 
\begin{center}
\vspace*{-2mm}
$\nu_{\rm br} \propto \frac{B}{(B^{2}+B_{\rm IC}^{2})^{2}} t^{-2}~~,
\hspace*{5mm}
\nu_{\rm br}^{\prime}=\nu_{\rm br} \cdot (1+t_{\rm CI}/t_{\rm RE})^{2}$~~,~~and~~~
\hspace*{0mm}
$\nu_{\rm br}^{\prime \prime}\propto \frac{B}{B_{\rm IC}^{4}} t^{-2}~~, $
\\[1.5ex]
\end{center}

\noindent
where $\nu_{\rm br}$ is the  break frequency of the {\it first} 
electron population injected at the beginning of the CI phase, and
$\nu_{\rm br}^{\prime}$ is the break frequency of the {\it last} electron
population injected just before the switch-off. 
The positions of $ \nu_{\rm br}$ and $ \nu_{\rm br}^{\prime}$ 
allow one to determine the durations of the CI and RE phases. The third 
high-frequency break, $ \nu_{\rm br}^{\prime \prime}$, is due to the 
inverse-Compton losses of the small pitch-angle electrons. 
Since $B_{\rm IC}$ is known, the ratio between 
$ \nu_{\rm br}^{\prime \prime}$ and $\nu_{\rm br}$ allows
one  to estimate the magnetic field strength directly.

In the emission spectrum, three asymptotic  power laws  
($S_{\nu} \propto (\nu/\nu_{\circ})^{-\alpha}$) can be identified:
\begin{itemize}
\item[a)]$\nu \ll \nu_{\rm br}$, the spectral index is $\alpha=\alpha_{\rm inj},$
\item[b)]$\nu_{\rm br} \ll \nu \ll \nu_{\rm br}^{\prime}$, 
  the spectral index is $\alpha=\alpha_{\rm inj}+0.5,$
\item[c)]$\nu_{\rm br}^{\prime}\ll \nu \ll \nu_{\rm br}^{\prime\prime}$, 
  the spectral index is $\alpha=4/3\cdot\alpha_{\rm inj}+1.$
\end{itemize}
For  $\nu \gg \nu_{\rm br}^{\prime\prime}$, the spectrum cuts off 
exponentially.  Power law b) quickly disappears after the switch-off 
($t_{\rm RE} > 0.05 \cdot t_{\rm CI}$) and power law c) is seen clearly 
only if $B > 30\,B_{\rm IC}$.

The KGJP model is characterized by four free parameters:
\begin{itemize}
\item[1)] the injection spectral index of the CI phase, 
  $\alpha_{\rm inj}$,
\item[2)] the total source age, $t=t_{\rm RE}$+$t_{\rm CI}$,
\item[3)] the relative duration of the relic phase with respect to 
  the CI phase, $t_{\rm RE}$/$t_{\rm CI}$,
\item[4)] the flux normalization.  
\end{itemize}
In this case, the isotropy of the electron population does not 
permit one to measure the source magnetic field.
In the emission spectrum only two break frequencies appear: 
\begin{center}
\vspace*{-2mm}
$ \nu_{\rm br} \propto \frac{B}{(2/3\cdot B^{2}+B_{\rm IC}^{2})^{2}}~t^{-2} $, ~~and~~~~
$ \nu_{\rm br}^{\prime} = \nu_{\rm br} \cdot (1+t_{\rm CI}/t_{\rm RE})^{2} $. \\[1.5ex]
\end{center}
The asymptotic power-law regions are the same as the KGKP model except
for power law c) which never appears.  For $\nu \gg \nu_{\rm
br}^{\prime}$, the spectrum cuts off exponentially.  In the limit $B
\rightarrow B_{\rm IC}$ (for which inverse-
Compton losses are as important as synchrotron losses) any pitch-angle
anisotropy is canceled and the KGKP and the KPJP model converge to the
same spectral shape.

Komissarov \& Gubanov (1994) concluded that the isotropic KGKP 
model provides a better fit to the data for their relic sample. 
In this way they were able to estimate the magnetic field  
strength of the relics by radio observations alone.
We have added to the radio spectra of the relics in 
A13, A85, A133 and A4038, as compiled from the literature,
the new flux-density measurements at 1.425\,GHz, and we have fitted 
the KG models using the program Synage++ (Murgia 2001, in preparation).
The fitting procedure finds the best free parameters 
and the confidence regions using a minimum $\chi^{2}$ technique.
Following Komissarov \& Gubanov, we kept the injection
spectral index fixed at 0.5. We confirm that the KGKP is the best model
for all the relics, although for A85 both the KGKP and the 
KGJP models give comparable $\chi^{2}$ values.
The best-fit parameters and the physical parameters obtained 
for the KGKP are summarized in Tables~4 and 5, respectively. 
The model fits are shown in Fig.~8, along with the  CI
spectra expected at the end of the active phase (see \S\ref{lumprog}).
The radiative ages have been calculated according to:

\begin{equation}
t=1060\,\frac{B^{0.5}}{B^{2}+B_{\rm IC}^{2}} ((1+z)\cdot \nu_{\rm br})^{-0.5}~, \nonumber
\end{equation}

\noindent
where the break frequency is expressed in GHz,  the magnetic field
in $\mu$G and the radiative age in Myr.  In the age calculation the magnetic fields resulting from
the fits have been used. These are consistent with the magnetic fields
derived by the minimum-energy assumption, $B_{\rm me}$, for A85 and
A133 and similar values were derived by Komissarov \& Gubanov (1994)
for the KP model. For the cases of A13 and A4038, $B_{\rm me}$ is
weaker by factors of 3.8 and 1.7, respectively.

The minimum-energy magnetic field was calculated using 
Miley (1980) Eq.~(2) and the standard assumptions that the lower and upper
cut-off frequencies are 10\,MHz and 100\,GHz, the ratio of energy in
heavy particles to that in the electrons is unity, the filling factor
in the emitting regions is unity, the angle between the magnetic field
and the line of sight is 90\deg.  We assumed the depth of the source,
$s$, to be the linear distance corresponding to the FWHM beam
diameter, $\sqrt{\theta_x\,\theta_y}$, where $\theta_x$ and $\theta_y$
are the major and minor axes of the restoring beam.  The spectrum that
is integrated by Miley in Eq.~(2) is a single power law with spectral
index $\alpha$.  However, the observed relic spectra are strongly
curved and are not well represented by a single power law.  Instead,
we treated the spectra initially as a series of piece-wise power laws
between the spectral measurements in Table~3 and applied Eq.~(2) of
Miley to integrate the spectrum over each frequency interval.  Then,
to extrapolate to the full span of 10\,MHz to 100\,GHz using Eq.~(2)
of Miley, we found an ``equivalent spectral index'' that yielded the
same result as the piece-wise integration when integrated over the
same range of frequencies.  For the integration of the spectrum from
10\,MHz to 100\,GHz this equivalent spectral index was then used in
Eq.~(2) of Miley, along with $F_0/(\theta_x \theta_y)$, where $F_0$ is
the median surface brightness of the relic at 1.4\,GHz.  This
procedure yielded the minimum-energy magnetic field strengths listed
in Tables~5 and 7.  The median $B_{\rm me} = 8.8$\,$\mu$G is
significantly lower than that derived by Slee \& Reynolds (1984) for
six steep-spectrum sources (including the present four relics) of 
$B_{\rm me} = 15.5$\,$\mu$G.  The difference can be attributed to our 
weighting of the spectral indices to account for spectral curvature.

\subsection{The Diffusion MJP Model}  \label{mjpmodel}

Here, we develop the KG models by including the effects of particle
diffusion between regions of different magnetic field strength, to
model the effect on the spectra of the filamentary structure.

An important property of the relic spectra is that the high-frequency
steepening is less than exponential, which is best fit by the
KGKP-based model when one assumes that the magnetic field is uniform
and constant, and hence that there is no pitch angle
scattering. However, the KGJP model is considered to be more realistic
from a physical point of view, since an anisotropic pitch angle
distribution (KP) will self-induce Alfv{\'e}n waves that will scatter
the electron in pitch angle (see the discussion by Carilli et al.\
1991). Even if this does not occur, the pitch angle will change when
the particle moves between regions of different magnetic field
strength.

The detailed images of the relics presented in this work show a variety of 
fine structure in the form of diffuse emission and filaments. If the
brightness contrast between these regions is due to a variation of the
magnetic field strength then the superposition of different isotropic spectra,
with a range of shifted break frequencies, can mimic the behavior of the 
anisotropic KGKP model. The effect is particularly noticeable for integrated 
spectra since all the regions of the source are seen together.     

Following Tribble (1993), we explore this alternative by supposing
that pitch-angle scattering occurs along with diffusion in a spatially
variable magnetic field. We approximate the magnetic field strength
distribution with a Maxwellian distribution characterized only by its
mean value $\langle B \rangle$. The radiative losses suffered by a
single electron are a combination of local losses associated with the
magnetic field where the electron is radiating now, and global losses
associated with the average field seen by the electron during its
diffusion and with the inverse-Compton scattering. Following Tribble
(1993), we introduce a ``diffusion efficiency'' $D_{\rm eff}$. The
electron energy losses due to synchrotron radiation are proportional
to a combination of the local field, $B$, and the average field:
\begin{equation}
  D_{\rm eff} \langle B^2\rangle + (1-D_{\rm eff}) 
\cdot B^2
\end{equation}
$D_{\rm eff}$ and $(1-D_{\rm eff})$ are the fractions of a particle's 
lifetime that it has spent in the average field $\langle B \rangle$ 
and in the actual field, $B$, respectively.
For simplicity we consider the  diffusion efficiency coefficient to be 
independent of the energy and the age of the electrons.
In the extreme case $D_{\rm eff}=0$ (no diffusion) the losses are 
determined only by the local field.  The break frequency of a
particular population of electrons with the same age varies 
as $\nu_{\rm br} \propto B^{-3}$.
When we integrate the contributions from all the 
different field strength regions the sharp JP cut-off is 
smoothed out and we see a KP-like spectrum.
The other extreme corresponds to $D_{\rm eff}=1$ (strong diffusion).
In this limit the electron energy spectrum is the same everywhere 
because the electrons have sampled all the field strengths. 
The break frequency varies from place to place as 
$\nu_{\rm br} \propto B^{-1}$. When we integrate the contributions from 
all the different field strength regions the sharp exponential JP 
cut-off is only partially dimmed and we still see a JP-like spectrum. 
Since $B_{\rm IC}$ acts as a global field, in the presence of 
inverse-Compton losses  we can state a lower limit for the 
diffusion efficiency\,:
\begin{equation}
 D_{\rm eff} \ge \frac {1} {1+ \langle B^{2}\rangle / B_{\rm IC}^{2} } \nonumber
\end{equation}
This offers the opportunity to put a lower bound on the source 
magnetic field from the spectral fits.\\

The MJP model (Murgia 2001, in preparation) extends Tribble's (1993)
analysis to a situation in which:  
\begin{itemize}
\item[a)] there is an injection phase (CI) of particles with a power law 
  energy spectrum and then a switch-off followed by a relic phase,

\item[b)] the pitch angle distribution is isotropic (JP),

\item[c)] the magnetic field is inhomogeneous (i.e.\ it has fine structure),

\item[d)] particles may diffuse between regions of different
  magnetic field strength.
\end{itemize}
There are five free parameters in the MJP model:
\begin{itemize}
\item[1)] the injection spectral index of the CI phase, $\alpha_{\rm inj}$,
\item[2)] the total source age, $t=t_{\rm RE}$+$t_{\rm CI}$,
\item[3)] the relative duration of the relic phase with respect to the CI phase, $t_{\rm RE}$/$t_{\rm CI}$,
\item[4)] the diffusion efficiency, $D_{\rm eff}$,
\item[5)] the flux normalization.  
\end{itemize}
The best-fit parameters and the physical parameters obtained for the MJP 
are summarized in Tables~6 and 7, respectively. The model fits are shown 
in Fig.~9, along with the  CI spectra expected at the end of the active 
phase (see \S\ref{lumprog}).
The goodness of the fits obtained for the MJP model is comparable to that 
of the KGKP for A85 and A4038, and slightly worse for A13 and A133. The limits 
inferred for the magnetic field strength are consistent with those of the KGKP.
The radiative ages from the MJP model have been calculated according to:
\begin{equation}
t=1060\,\frac{\langle B \rangle ^{0.5}}
{2/3 \langle B^{2} \rangle 
  +B_{\rm IC}^{2}} ((1+z)\cdot \nu_{br})^{-0.5}~, \nonumber
\end{equation}

\noindent
where the break frequency is expressed in GHz, the magnetic field in $\mu$G and 
 the radiative age in Myr.
We have assumed that $\langle B \rangle$ is the minimum-energy 
magnetic field ($B_{\rm me}$), since the field is not computed 
from the MJP model.  The lower limits on the mean magnetic field 
strength obtained from the estimate of the diffusion efficiency  
are consistent with those estimated from the minimum-energy 
assumption for A85, A133 and marginally for A4038 also. In the 
case of A13, $B_{\rm me}$ is a factor of 3.7 weaker with respect 
to the lower limit estimated from the fit.

\subsection{Luminosities of Progenitors} \label{lumprog}

Both the KGKP and MJP models describe a scenario in which adiabatic
expansion and/or compression events are neglected, the magnetic field
intensity does not evolve, there is no substantial particle
reacceleration and the relativistic electrons do not escape from the
source. Indeed, only the radiative losses are supposed to influence
the shape of the relic's spectra after the switch-off.  Under these
assumptions we extrapolated the relics' spectra back to the end of the
continuous injection phase.  These are the CI models shown in Figs.~8
and 9. By measuring the difference between the observed flux densities
and those predicted for the CI model we have calculated the 1.4
GHz\,radio power expected for the progenitor sources at the
switch-off instant (see Tables~5 and 7). We found that
$P_{\mathrm{1.425\,GHz}}$ ranges from $10^{24.4}$~W\,Hz$^{-1}$ to
$10^{25.5}$~W\,Hz$^{-1}$ and from $10^{25.1}$~W\,Hz$^{-1}$ to
$10^{26.1}$~W\,Hz$^{-1}$ for the KGKP and MPJ model, respectively. The
corresponding total luminosities range from 10$^{34}$\,W to
10$^{36}$\,W.  These relatively high $P_{\mathrm{1.425\,GHz}}$ values
place the progenitors on the boundary between FR\,I and FR\,II radio
galaxies, and hence among the most luminous 7\,\% of radio galaxies
(e.g. Slee et al.\ 1998).  Therefore, it seems that only the most powerful
radio galaxies in clusters have left radio relics that are detectable
with the present generation of synthesis radio telescopes and explains
the paucity of detected relics.

\subsection{Summary of Spectral Modelling Results}  \label{sumanal}
 
We confirmed that the extremely steep spectra of the relics studied in
this work are consistent with the scenario in which the injection of
fresh electrons in these sources has ceased for a significant fraction
of their lifetime. However, the flux-density measurements indicate
that the spectral cut-offs are less than exponential.  Under the
simple hypothesis that the magnetic field in the relics is uniform, we
found that the anisotropic KGKP model fits better than the isotropic
KGJP model.  In contrast, our VLA images show that the structure of
the relics is quite inhomogeneous.  By fitting the MJP model we
demonstrated that an isotropic distribution of electrons can still
reproduce the observed spectra if the brightness contrasts are due to
variations of the magnetic field strength.  The goodness of the MJP
fits is comparable to that of the KGKP, although the KGKP is always
slightly better.

The most convincing results were found for A85 and A133 because of the
good frequency coverage of their integrated spectra.
For these two relics, both the value of the magnetic field strength
obtained from the KGKP fits and the lower limits obtained from the MJP are
fully consistent with the magnetic fields estimated by minimum-energy
arguments.

The integrated spectra of A13 and A4038 were not so well
defined and did not allow us to constrain all the fit parameters
(particularly for A13).
In these cases, the estimates of the magnetic field strengths obtained 
from fits differ significantly from $B_{\rm me}$.
The radiative ages derived from the spectral break frequencies 
range from $0.3\,\times\,10^{8}$\,yr to  $2.3\,\times\,10^{8}$\,yr for the 
KGKP model and from $1.0\,\times\,10^{8}$\,yr to $2.4\,\times\,10^{8}$\,yr 
for the MJP model. 

Finally, we compared the model spectral index at 1.4\,GHz resulting from
the fit with the spectral index measured from our VLA 1.4 GHz images (Table~2).
The model spectral indices for A13, A85, A133 and A4038 are 
2.1, 3.8, 2.5, and 2.6, and the observed spectral indices
are 4.4, 3.0, 2.1, and 3.1 respectively.  
Both the KGKP and the MJP models predict very similar spectral indices
at 1.4\,GHz.  

The observed spectral indices were underestimated by the
models for A13 and A4038, and overestimated for A85. The spectral
index of A133 is fairly well predicted.  The most serious discrepancy
is found for A13, and it is problematic to explain the origin of
this mismatch.  In fact, independent of any model, the integrated
spectrum of A13 between 160\,MHz and 1.4\,GHz is almost a power law
with a spectral index of $\alpha \sim 2$.  The sudden steepening of
the spectrum measured from the VLA images is difficult to reconcile
with the trend of the integrated spectrum at lower frequencies.  The
discrepancy could be observational or a real
effect.  The observed spectral index from the high-resolution VLA
measurement could be overestimated due to missing flux from a more
extended low-brightness component. However, this seems
unlikely because Table~2 shows that our flux-density measurement with
the B array actually exceeds that of the NVSS with the D array. Maybe
we are witnessing the onset of an exponentially steepening spectral 
index above 1\,GHz as predicted by the JP model.  The same
process is also present to a lesser degree in the relics of A85 and
A4038.

\vspace*{3mm}

\section{Discussion}  \label{discussion}

\subsection{Relic Origins and Spectral Models} \label{disc_origins}

The radio relics observed here show a surprising amount of fine
structure at $4''$ resolution.  This required us to generalize the
spectral-ageing models of relics to include the diffusion of particles
between regions of non-uniform magnetic field (\S\ref{mjpmodel}),
which yielded better success in fitting the relic spectra
than did the scenario for the production of halo and edge relics
that used reacceleration of seed relativistic electrons by shocks produced
during cluster mergers.  Although Roettiger, Burns and Stone (1999) 
had good success modelling
the edge relics in A3667 with this latter approach, 
the reacceleration process cannot reproduce the much more extreme spectral
indices nor the fine structure of the relics considered here.  Recent
work by En{\ss}lin \& Gopal-Krishna (2001) models the relic in A85 as a
recompressed radio lobe that had previously expanded and cooled
adiabatically, and they obtain an equally good fit to the spectra as we
achieved.  We shall probably be able to
distinguish between the competing mechanisms of an ageing relativistic
population and ageing combined with recompression, when more accurate
spectral and polarization data are available.

The referee noted a similarity between our relics and the tailed radio
galaxy in A3556 which has lost its source of fresh relativistic
electrons and has the former host galaxy situated near one extremity
of the tail (Venturi et al.\ 1998).  This seems reasonable, however we
note that the radio structures are rather different and that still needs to 
be explained.  The tailed radio galaxy is predominantly a
linear structure whereas the relics that we have imaged are more
complex.  The SE extension of the relic in A13 looks like a tailed
remnant whose host would be galaxy $C$ or $E$ in Fig.~2, but the
remainder of the structure is not linked directly to any nearby
galaxy.  In A85 and A4038, the likely host galaxies are much more
distant from the radio structures than in A3556.  
The spectra of the relics are also different from the A3556 source,
all four being much steeper.  This can be explained naturally if our
relics have aged longer than has the A3556 source.
To compare the spectral ages, we have reanalysed the flux-density measurements
of A3556 by Venturi et al.\ (1998) using the KGKP and MJP models and found
the best fit with the MJP model.  This model infers a
mean field strength $<B>\,\simeq\,1.7\,\mu$G, which agrees well with the
equipartition field strength of 1.6\,$\mu$G calculated by Venturi et al.,
and it infers a high value for the diffusion efficiency ($D_{\rm eff} = 0.75$).
Since the spectrum of the A3556 source is not as steep as our relics,
the duration of the relic phase in A3556, $t_{\rm RE}$, should be short 
compared to the active phase, $t_{\rm CI}$.  In fact, the model fits found
$t_{\rm RE} / t_{\rm CI} = 0.1$ using the KGKP model and 
$t_{\rm RE} / t_{\rm CI} = 0.07$ using the MJP model.
These values are considerably lower than those found for
our relics and confirm the recent formation of the A3556 tail 
as a relic. The ages found for the A3556 relic were (KGKP model) 
$t = 520$\,Myr, $t_{\rm CI} = 470$\,Myr and $t_{\rm RE} = 50$\,Myr, and
using the MJP model: $t = 490$\,Myr, $t_{\rm CI} = 458$\,Myr and 
$t_{\rm RE} = 32$\,Myr.  The duration age of the relic phase, $t_{\rm RE}$
is comparable to those found for our relics, but more important for
the spectral steepening is that the relative value of
$t_{\rm RE} / t_{\rm CI}$ is smaller for the A3556 source.

\subsection{Optical Identifications} \label{disc_optids}

The identification of former host galaxies for our radio relics is
problematic (\S\ref{optids}). The identification of the relics with
the BCGs is by no means certain, because of the disagreement between
the modelled relic ages and the galaxy travel times.  In fact, there
is usually at least one other relatively bright galaxy near the relic
that provides a closer match between relic age and travel time than 
does the BCG, and we found that the most likely identification was with 
those rather than the BCG for two of the four relics.

\subsection{Pressure Balance} \label{disc_pressure}

We can compare the pressure of the relativistic plasma, $p_{\rm r}$,
derived from the radio emission to the pressure of the intra-cluster
gas, $p_{\rm ic}$, derived from its X-ray properties, and so
understand something about the confinement of the relic.  We assumed
approximate equipartition between field and particle energies in the
radio source, and static isothermal pressure due to the intra-cluster
gas.  There is not enough information in existing X-ray images to determine
radial temperature and electron-density profiles for A13, A133 and A4038,
and so we have adopted the profiles determined for A85 by 
Pislar et al.\ (1997) and assumed that they are the same in 
A13, A133 and A4038.

The pressure of the relativistic plasma is $\mu_{\rm me}/3$, where 
$\mu_{\rm me}$, the minimum energy density, was derived from Miley 
(1980) Eq.(1).  The thermal pressure of the intra-cluster gas
is $n\,k\,T\,/\,A_{\rm r}$ dyne\,cm$^{-2}$,
where $n$ is the
electron density in cm$^{-3}$, $T$ the temperature in K, 
$k = $1.38$\times$10$^{-16}$\,erg\,K$^{-1}$ and $A_{\rm r}$ = 0.62 is the 
relative atomic mass for a fully ionized gas containing 10\,\% He
by number. The temperature is scaled from  $k\,T\,=\,$4\,keV 
(Pislar et al.\ 1997) for A85, with the assumption that 
$T$\,$\propto$\,(velocity dispersion)$^2$ (Mazure et al.\ 1996). 
The central electron density of each cluster is scaled from that of
A85 on the assumption that it is proportional to the square root of the
X-ray luminosity.
Therefore, this assumes that the clusters have the same core radius and
radial variation in electron density as A85. Model No.~2 from 
Pislar et al.\ (1997) was then used to evaluate the electron densities, $n$,
at the centroid of the relic, and are given in Table 8.

Comparing the relativistic plasma pressure to the thermal pressure of
the intra-cluster gas, we find that pressure equilibrium exists only
for the relic in A85. In the remaining three relics, the pressure of
the thermal gas is three to five times that of the relativistic
plasma.  However, the value of $B_{\mathrm{me}}$ in Table 8 could well
be increased enough by relatively small departures in the assumptions
used to calculate $B_{\mathrm{me}}$ (the square of which determines
$p_{\mathrm{r}}$) to bring $p_{\mathrm{r}}$ and $p_{\mathrm{ic}}$ to
near equality.  We conclude that the radio plasma and intra-cluster
thermal gas are approximately in pressure equilibrium, allowing the
relic to age and retain its identity for at least $10^8$ yr.

The relics could not be significantly over-pressured or they
would have expanded and faded to undetectable levels in $10^8$ yr.
For adiabatic expansion of a relativistic magnetised
plasmon, the total synchrotron emission scales as $r^{-10.4}$ for
$\alpha_{\rm 1.425\,GHz} = 2.1$ as in the A133 relic (e.g. Lang 1999).
Expansion by a factor of only 1.6 from the present radius would reduce
the flux density of our strongest relic A133, for example, from the
present 168 mJy down to the 1 mJy flux-density limit of the original
Slee et al. finding survey.  
Since this has not happened, the relics are probably
confined by the external pressure, which also suggests
that the standard assumptions used for calculating the relativistic 
plasma pressure are not too far from reality.

However, En{\ss}lin \&
Gopal-Krishna (2001) propose that such expansion and cooling does
actually occur and that relics fade and lurk, unseen, for long periods 
until they are later recompressed by shocks during subsequent cluster mergers
and light up once more.

\subsection{X-ray Emission Mechanisms} \label{disc_ic}

We found a suggestion of excess X-ray emission from the region of the
radio relics in Figs.~11, 12 and 13.  An excess is expected due to
inverse-Compton scattering of CMB photons.  However, the expected
brightness, using the equipartition magnetic field strengths, should
be much less than that of the observed excesses, and so the excess is
probably due to other emission mechanisms.
This situation has been considered in
some detail in the case of the Coma halo by En{\ss}lin et al. (1999),
Dogiel (2000), and Liang (2001) who find that the high-energy excess
is likely to be bremsstrahlung from a non-thermal tail of the electron
energy distribution, and Liang found that the inverse-Compton
component should be weaker than the bremsstrahlung by orders of
magnitude.  If such considerations apply to the present relics, 
the excess emission might be bremsstrahlung from sub-structure in 
the ICM or from a background cluster projected onto the relic.

Bagchi, Pislar and Lima Neto (1998) have already suggested that the
excess X-ray emission in A85 (see Fig.~11) is inverse-Compton
scattering and derive a field of about 1\,$\mu$G.  Following their
approach and using our data on A85 we derive magnetic field strengths
of $0.47$\,$\mu$G for A85 and $0.60$\,$\mu$G for A133.  If the
identification of the excess X-rays with inverse-Compton emission is
correct, then the factors of 18 to 24 difference between the
fields and the presently deduced $B$ and $B_{\rm me}$ from the
spectral modelling in \S\ref{specanal} has three implications; 1)
there is a gross mismatch between field and particle energy densities;
2) the mismatch would invalidate the present modelling; 3) the relic
would be under-pressured to such an extent that it could not maintain
its identity against the pressure of the hot intra-cluster gas.

Alternatively, the difference between inverse-Compton and
minimum-energy field strengths could be explained if the
electrons that produce inverse-Compton emission stay in regions of low
magnetic field strength and thus would not be visible in a
high-freqeuncy radio image.  The relic radio emission would be
tracing only the peaks of the magnetic field strength distribution.

An inverse-Compton component is likely to be identified only in a
higher energy range than that covered by ROSAT, where better contrast
between the thermal cluster emission and inverse-Compton emission is
expected.  Such observations require enough angular resolution to see
features that match the structures in our radio images and enough spectral
resolution to distinguish between the various emission processes.
Such experiments may be possible with the present generation of X-ray 
telescopes such as {\it XMM-Newton} and {\it Chandra}.

\acknowledgments
The National Radio Astronomy Observatory is a facility of the National Science
Foundation operated under cooperative agreement by Associated Universities, Inc.
The ROSAT project is supported by the German Bundesministerium f\"ur
Bildung und Forschung (BMBF/DLR) and the Max-Planck-Gesellschaft (MPG).
This research has made use of the NASA/IPAC Extragalactic Database (NED)   
which is operated by the Jet Propulsion Laboratory, California Institute   
of Technology, under contract with the National Aeronautics and Space      
Administration.
The Digitized Sky Survey was produced at the Space Telescope Science
Institute under U.S.\ Government grant NAG~W-2166. The images of these
surveys are based on photographic data obtained using the Oschin Schmidt
Telescope on Palomar Mountain and the UK Schmidt Telescope. The plates were
processed into the present compressed digital form with the permission of
these institutions.
HA thanks CONACyT for financial support under grant 27602-E.
MM acknowledges a partial support by the Italian Ministry for University
and Research (MURST) under grant Cofin98-02-32.
We thank the referee for his valuable critique and suggestions for
improvement.

\scriptsize
\begin{deluxetable}{lccccccccc}
\tablenum{1}
\tablecolumns{10}
\tablecaption{Parameters of Observed Clusters} \label{tab1}
\tablehead{
\colhead{Cluster}&
\colhead{R.A.~~Dec.}&
\colhead{$R$~~$BM$}&
\colhead{$<z_{\mathrm{hel}}>$}&
\colhead{$\sigma_{\rm V}$}&
\colhead{$R_{\mathrm{A}}$}&
\colhead{lg\,$(L_{\mathrm{X}}$}&
\colhead{$\sigma$} &
\colhead{LLS} \nl
\colhead{Field}&
\colhead{(J2000)} &
\colhead{} &
\colhead{} &
\colhead{/ km\,s$^{-1}$}&
\colhead{(\,$'$\,)}&
\colhead{/ erg\,s$^{-1}$)}&
\colhead{/ $\mu$Jy beam$^{-1}$} &
\colhead{(kpc)} \nl
\colhead{(1)} &\colhead{(2)} &\colhead{(3)~~(4)} &\colhead{(5)} & \colhead{(6)} 
  &\colhead{(7)} &\colhead{(8)} &\colhead{(9)} &\colhead{(10)} 
}
\startdata
A~~13  & 00 13 38.5 $-$19 30 03 & ~2~\,III & 0.0943 \tablenotemark{a} & 886 & 21.3 & 44.2 & 17 & 
   261 \nl
A~~85  & 00 41 50.5 $-$09 18 12 & 1~~I   & 0.0555 \tablenotemark{b} &   957 & 34.0 & 44.8 & 17 & 
  150 \nl
A~133  & 01 02 41.6 $-$21 52 55 & 1~~I   & 0.0562 \tablenotemark{c} &  735 & 33.5 & 44.4 & 21 & 
   56 \nl
A4038  & 23 47 45.1 $-$28 08 26 & 2~~I   & 0.0292 \tablenotemark{a} &  839 & 63.9 & 44.1 & 20 & 
   56 \nl
\enddata

\tablecomments{Col 2: Coordinates of the brightest cluster galaxy, as
determined from the Digitized Sky Survey.  Cols 3 \& 4: Abell richness
class ($R$), and Bautz-Morgan morphological type ($BM$) 
from Abell et al.\ (1989).  Col 5: $<z_{\mathrm{hel}}>$ is the mean
heliocentric redshift of the cluster, from (a) \cite{mazure96}, (b)
\cite{durret98}, and (c) \cite{way97}.  Col 6: $\sigma_{\rm V}$ is the
cluster velocity dispersion from the same references.  Col.~(7) gives
the Abell radius ($R_{\mathrm{A}} = 2$\,Mpc at the cluster distance).
Col 8: $L_{\mathrm{X}}$ is the X-ray luminosity (in
erg\,s$^{-1}$, for the band 0.1\,keV to 2.4\,keV) from \cite{ebeling96}; 
Col.\ (9) gives the noise of our radio images, in $\mu$Jy\,beam$^{-1}$.
Col.\ (10) gives the largest linear size of the radio relic in kpc,
assuming cluster membership.}

\end{deluxetable}

\begin{deluxetable}{lccccc}
\tablenum{2}
\tablecolumns{6} 
\tablecaption{Integrated Radio Parameters} \label{tab2}
\tablehead{
\colhead{Relic} &
\colhead{$S_{\mathrm{1.425\,GHz}}$} &
\colhead{lg ($P_{\mathrm{1.425\,GHz}}$} &
\colhead{Flux Deficit} &
\colhead{Spectral Index} &
\colhead{Polarization} \nl
\colhead{Source} & 
\colhead{/ mJy} & 
\colhead{/ W\,Hz$^{-1}$)} & 
\colhead{/ mJy} & 
\colhead{at 1.425\,GHz} &
\colhead{(\%)} \nl
\colhead{(1)} &\colhead{(2)} &\colhead{(3)} &\colhead{(4)} & \colhead{(5)} &\colhead{(6)} 
}
\startdata
A13\_6a/b/c             & ~35.5 $\pm$ 1.7    & 23.63 & $-$7.4 $\pm$~2.3 & 4.4 $\pm$ 0.4  &   11.6 $\pm$ 4.1 \nl
A85\_25a/b/c            & ~40.9 $\pm$ 2.3    & 23.32 & $-$0.4 $\pm$~3.4 & 3.0 $\pm$ 0.2  &   16.4 $\pm$ 8.8 \nl
A133\_7a \tablenotemark{a} & 137.~ $\pm$ 6.~\, & 23.88 &  ~~8.1 $\pm$13.6 & 2.1 $\pm$ 0.1  &   ~2.3 $\pm$ 1.5 \nl
A133\_7b                & ~23.0 $\pm$ 1.1    & 23.10 &                  & 2.1 $\pm$ 0.1  &                  \nl
A133\_6                 & ~~4.9 $\pm$ 0.3    & 22.45 &                  & 1.4 $\pm$ 0.2  &                  \nl
A133\_8                 & ~26.5 $\pm$ 1.8    & 24.52 &                  & 1.5 $\pm$ 0.1  &                  \nl
A4038\_9 \tablenotemark{b} & ~49.0 $\pm$ 2.4 & 22.87 &  ~~8.6 $\pm$ 5.5 & 3.1 $\pm$ 0.1  &   ~4.6 $\pm$ 2.3 \nl
A4038\_10                & ~~2.3 $\pm$ 0.1   &       &                & $-$0.5 $\pm$ 0.2  &                  \nl
A4038\_11                & ~23.3 $\pm$ 1.0   & 22.58 &                  & 0.4 $\pm$ 0.1  &                  \nl
\enddata

\tablecomments{Col.~(1): Source names from Slee et al.\ (1994).
Col.~(2): 1.425\,GHz flux density from our observations.
Col.~(3): decimal logarithm of the 1.425 GHz radio power, assuming
  cluster membership for the sources.
Col.~(4): NVSS flux density minus 1.425\,GHz flux density in Col.~(2).
Col.~(5): Spectral index determined from the images at 1.385\,GHz and
1.465\,GHz.  The error estimate is based on the random noise level in
each image and does not include systematic errors.  See \S\ref{highres} 
for further explanation.
Col.~(6): integrated fractional linear polarization at 1.425\,MHz.
The standard deviation quoted reflects the variation in different
parts of the relic.  See \S\ref{polari}.}
\tablenotetext{a}{~Source 7a is the relic; 7b is the cD galaxy; sources 6 
(a cluster galaxy) and 8 (a background galaxy) are not properly 
resolved in lower-resolution measurements.}
\tablenotetext{b}{~Source 9 is the relic; 10 (unidentified) and 11 (the cD)
are sources not properly resolved in lower-resolution measurements.}
\end{deluxetable}

\tiny
\begin{deluxetable}{cccccccccc}
\tablenum{3}
\tablecolumns{10}
\tablecaption{Radio Flux Densities of the Four Relic Sources} \label{tab3}
\tablehead{
\colhead{Freq.} &
\colhead{Abell 13}  &
\multicolumn{2}{c}{Abell~85} &
\multicolumn{2}{c}{Abell~133~\tablenotemark{b}} &
\multicolumn{2}{c}{Abell~4038} &
\colhead{Ref.~\tablenotemark{c}} &
\colhead{Notes~\tablenotemark{d}} \nl
\colhead{(MHz)} &
\colhead{$S_{\mathrm{corr}}$ / Jy~\tablenotemark{a}}  &
\colhead{$\Delta$\,S / Jy} &
\colhead{$S_{\mathrm{corr}}$ / Jy}  &
\colhead{$\Delta$\,S / Jy} &
\colhead{$S_{\mathrm{corr}}$ / Jy}  &
\colhead{$\Delta$\,S / Jy} &
\colhead{$S_{\mathrm{corr}}$ / Jy}  &
\colhead{} &
\colhead{} \nl
\colhead{(1)} &\colhead{(2)} &\colhead{(3)} &\colhead{(4)} &\colhead{(5)} &\colhead{(6)}
   &\colhead{(7)} &\colhead{(8)} &\colhead{(9)} &\colhead{(10)} 
}
\startdata
16.7 &               & 45 $\pm$ 21 & 93   $\pm$ 24  &                &                  &             &                      & 1 & 1 \nl
29.9 &               & 23 $\pm$ ~9 & 93   $\pm$ 13  &   13 $\pm$ 5   & 46 $\pm$ 13      &  10 $\pm$ 4 & 32 $\pm$ 7           & 2 & 2 \nl
80~~ & 6.0   $\pm$ 1.2  &        & 34.0 $\pm$ 3.7   & 0.5 $\pm$ 0.2  & 35.5 $\pm$ 4.3   &             &  19.0 $\pm$ 2.7      & 3 & 3 \nl
160~~& 2.8   $\pm$ 0.6  &        &  8.33 $\pm$ 0.7  &                & 10.9  $\pm$ 1.2  &             &  4.3 $\pm$ 0.5       & 3 & 3 \nl
327~~& 0.63  $\pm$ 0.06 &        &  3.2 $\pm$ 0.32  &                &  2.82 $\pm$ 0.28 &  0.10 $\pm$ 0.02 & 1.44 $\pm$ 0.15 & 4 & 4 \nl
408~~& 0.49  $\pm$ 0.08 &        &  1.54 $\pm$ 0.25 & 0.04 $\pm$ 0.01& 2.62  $\pm$ 0.25 &  0.05 $\pm$ 0.01 &  0.91 $\pm$ 0.11 & 5,6 & 5 \nl
843~~& 0.09  $\pm$ 0.01 &        & 0.20 $\pm$ 0.03  &                &  0.53 $\pm$ 0.06 &  0.04 $\pm$ 0.01 &  0.17 $\pm$ 0.03  & 5,6 & 6 \nl
1400~~& 0.030 $\pm$ 0.003 &      & 0.043 $\pm$ 0.003 &               & 0.168 $\pm$ 0.006 & 0.025 $\pm$ 0.003 & 0.061 $\pm$ 0.003 & 7 & 7 \nl
2700~~&                 &        & $<$0.01~~~~~~~~~  & 0.025 $\pm$ 0.002 & 0.029 $\pm$ 0.016 &             &             & 8,9 & 8 \nl
4900~~&                 &        &                  &                & 0.0040 $\pm$ 0.0003  &             &  $<$0.0014~~~~~  & 10 & 9 \nl
\enddata

\tablenotetext{a}{~The flux corrections, $\Delta$\,S, are subtracted from 
the observed values to obtain $S_{\mathrm{corr}}$.  There is no column with 
flux-density correction for A13 because there were no measurements  
that were contaminated by confusing sources that needed subtraction.}
\tablenotetext{b}{~Flux densities include the relic, the cD and A133\_6 (Slee et al.\ 1994).}
\tablenotetext{c}{~References\,:  
1.~Braude et al.\ (1981); 
2.~Finlay \& Jones (1973); 
3.~Slee (1995); 
4.~Joshi et al.\ (1986);
5.~Reynolds (1986); 
6.~Slee \& Reynolds (1984); 
7.~Condon et al.\ (1998);
8.~Andernach et al.\ (1986);
9.~Reuter \& Andernach (1990);
10.~Slee et al.\ (1994).
}
\tablenotetext{d}{~See notes on calibrations and flux-density corrections
  in the text of \S\ref{broadspec}.}
\end{deluxetable}

\normalsize
\begin{deluxetable}{lccc}
\tablenum{4}
\tablecolumns{4} 
\tablecaption{Relic Parameters Derived from Fitting the KGKP Model to the
Relic Radio Spectra.} \label{tab4}
\tablehead{
\colhead{Source} &
\colhead{$\nu_{\rm br}$~/~MHz} &
\colhead{$t_{\rm RE}/t_{\rm CI}$} &
\colhead{$B/B_{\rm IC}$}  \nl
\colhead{(1)} &\colhead{(2)} &\colhead{(3)} &\colhead{(4)} 
}
\startdata
A13       & 72  ($<$ 100)   &  6.1 (undefined)   & 6.7 (4.5--9)   \\
A85       & 27  (15--60)    &  0.5 (0.3--2)      & 2.3 (1.3--3.4) \\
A133      & 28  (15--70)    &  0.4 (0.23--0.9)   & 3.5 (3--4.1)   \\
A4038     & 39  (5--100)    &  0.9 ($>$ 0.2)     & 4.5 (2--5)     \\
\enddata

\tablecomments{Fit parameters are given with their 68\,\%-confidence levels 
(in brackets).}
\end{deluxetable}

\begin{deluxetable}{lcccccc}
\tablenum{5}
\tablecolumns{7} 
\tablecaption{Relic parameters derived from fitting the KGKP model to the 
relic radio spectra.} \label{tab5}
\tablehead{
\colhead{Source} &
\colhead{$t$} &
\colhead{$t_{\rm CI}$} &
\colhead{$t_{\rm RE}$} &
\colhead{$B$} &
\colhead{$B_{\rm me}$} &
\colhead{lg\,($P_{\mathrm{1.4\,GHz}}$} \nl 
\colhead{} &
\colhead{/\,Myr}  &
\colhead{/\,Myr}  &
\colhead{/\,Myr}  &
\colhead{/\,$\mu G$} &
\colhead{/\,$\mu G$} &
\colhead{/ W\,Hz$^{-1}$)} \nl    
\colhead{(1)} &\colhead{(2)} &\colhead{(3)} &\colhead{(4)} &\colhead{(5)} &\colhead{(6)} &\colhead{(7)} 
}
\startdata
A13                &  38    &   5   & 33     & 21   &  6.8 &  25.5 \nl
A85                & 297    & 198   & 99     & 6.8  &  8.7 &  25.2 \nl
A133               & 171    & 122   & 49     & 10.4 & 14.4 &  25.1 \nl
A4038              & 112    &  59   & 53     & 12.6 &  8.9 &  24.4 \nl
\enddata

\tablecomments{Relic ages and magnetic fields are derived 
from the fits. The minimum-energy magnetic fields are also reported 
for comparison. The last column indicates the 1.4\,GHz radio power 
expected for the progenitor sources at the instant of switch-off.}

\end{deluxetable}

\begin{deluxetable}{lccc}
\tablenum{6}
\tablecolumns{4} 
\tablecaption{Relic Parameters Derived from Fitting the MJP Model to the
Relic Radio Spectra.} \label{tab6}
\tablehead{
\colhead{Source} & 
\colhead{$\nu_{\rm br}$/\,MHz} & 
\colhead{$t_{\rm RE}/t_{\rm CI}$} & 
\colhead{$D_{\rm eff}$} \nl
\colhead{(1)} &\colhead{(2)} &\colhead{(3)} &\colhead{(4)} 
} 
\startdata
A13       & 57  ($<$ 140)   &  9 (undefined)   &  0.02 (0.001--0.06)  \nl
A85       & 78  (27--135)   &  1 ($>$0.26)     &  0.24 (0.14--1)      \nl
A133      & 73  (27--129)   &  1 ($>$0.3)      &  0.08 (0.06--0.14)   \nl
A4038     & 101 (25--187)   &  3.9 (undefined) &  0.09 (0.05--0.27)   \nl
\enddata

\tablecomments{Fit parameters 
are given with their 68\%-confidence levels (in brackets).}
\end{deluxetable}

\begin{deluxetable}{lcccccc}
\tablenum{7}
\tablecolumns{7}
\tablecaption{Lower Limits to the Relic Ages and Magnetic Fields 
from the MJP Model.}  \label{tab7}
\tablehead{
\colhead{Source} &
\colhead{$t$} &
\colhead{$t_{\rm CI}$} &
\colhead{$t_{\rm RE}$} &
\colhead{$\langle B \rangle$} &
\colhead{$B_{\rm me}$} &
\colhead{lg\,($P_{\mathrm{1.4\,GHz}}$} \nl 
\colhead{} &
\colhead{/\,Myr}  &
\colhead{/\,Myr}  &
\colhead{/\,Myr}  &
\colhead{/\,$\mu G$} &
\colhead{/\,$\mu G$} &
\colhead{/\,W\,Hz$^{-1}$)} \nl    
\colhead{(1)} &\colhead{(2)} &\colhead{(3)} &\colhead{(4)} &\colhead{(5)} &\colhead{(6)} 
  &\colhead{(7)}
}
\startdata
A13               &  271    & 27    & 244      & $> 20$   & 6.8  &  26.1 \\
A85               &  184    & 92    & 92       & $>  5$   & 8.7  &  25.6 \\
A133              &   98    & 49    & 49       & $>  9$   & 14.4 &  25.5 \\
A4038             &  162    & 34    & 128      & $>  8$   & 8.9  &  25.1 \\
\enddata

\tablecomments{The minimum-energy magnetic fields used for the age
calculation are also reported. The last column indicates the 1.4\,GHz
radio power expected for the progenitor sources at the instant of
switch-off.}
\end{deluxetable}

\begin{deluxetable}{lcccccc}
\tablenum{8}
\tablecolumns{7}
\tablecaption{Relic Confinement Parameters} \label{tab8}
\tablehead{
\colhead{Cluster} &
\colhead{$B_{\rm me}$} &
\colhead{Relic Pressure} &
\colhead{Distance~\tablenotemark{a}} &
\colhead{$T_{\mathrm{cl}}$~\tablenotemark{b}} &
\colhead{$n_{\mathrm{e}}$~\tablenotemark{c}} &
\colhead{Thermal Pressure} \nl
\colhead{Field} &
\colhead{/\,$\mu G$} &
\colhead{/\,10$^{-12}$ dyne cm$^{-2}$} &
\colhead{/\,kpc} &
\colhead{/\,keV} &
\colhead{/\,10$^{-3}$ cm$^{-3}$} &
\colhead{/\,10$^{-12}$ dyne cm$^{-2}$} \nl
\colhead{(1)} &\colhead{(2)} &\colhead{(3)} &\colhead{(4)} &\colhead{(5)} &\colhead{(6)} &\colhead{(7)} 
}
\startdata
A13        &  6.8    &  1.4 & 137  & 3.4 & 0.5 &  4.2 \nl
A85        &  8.7    &  2.3 & 395  & 4.0 & 0.2 &  2.1 \nl
A133       & 14.4    &  6.4 &  36  & 2.4 & 2.8 & 25.1 \nl
A4038      &  8.9    &  2.4 &  47  & 3.1 & 1.5 & 13.5  \nl
\enddata
\tablenotetext{a}{~Projected distance between cluster and relic centroids.}
\tablenotetext{b}{~Cluster temperature (\S\ref{disc_pressure}).}
\tablenotetext{c}{~Electron density at the relic's centroid.}
\end{deluxetable}

\pagebreak

\begin{figure}
\epsfig{file=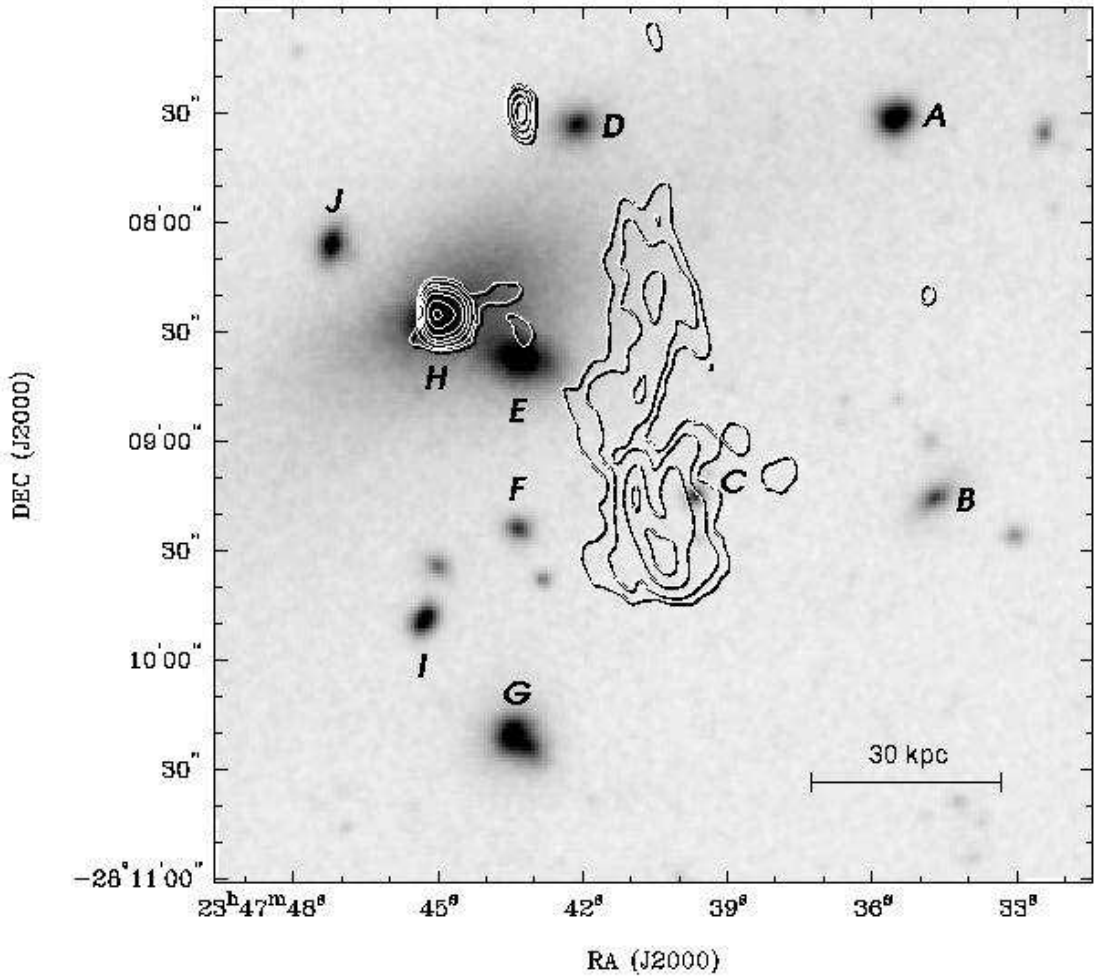,width=13cm,angle=-90}
\caption{
Radio contours of a naturally-weighted image
of the relic source in A4038 are overlaid on a red DSS-2 image of the area. 
The contour levels are (128, 256, 642, 1284)\,$\mu$Jy\,beam$^{-1}$ 
or $(4.1, 8.1, 20.3, 40.5)$\,K.  
The FWHM restoring beam for the radio image is 
$6.9'' \times 4.0''$ in PA$\,=\,2^{\circ}$.
The labelled galaxies are discussed in \S\ref{optids}.
The horizontal bar indicates the
linear scale, using the redshift from Table~1.
\label{fig7}}
\end{figure}

\begin{figure}
\mbox{
\epsfig{file=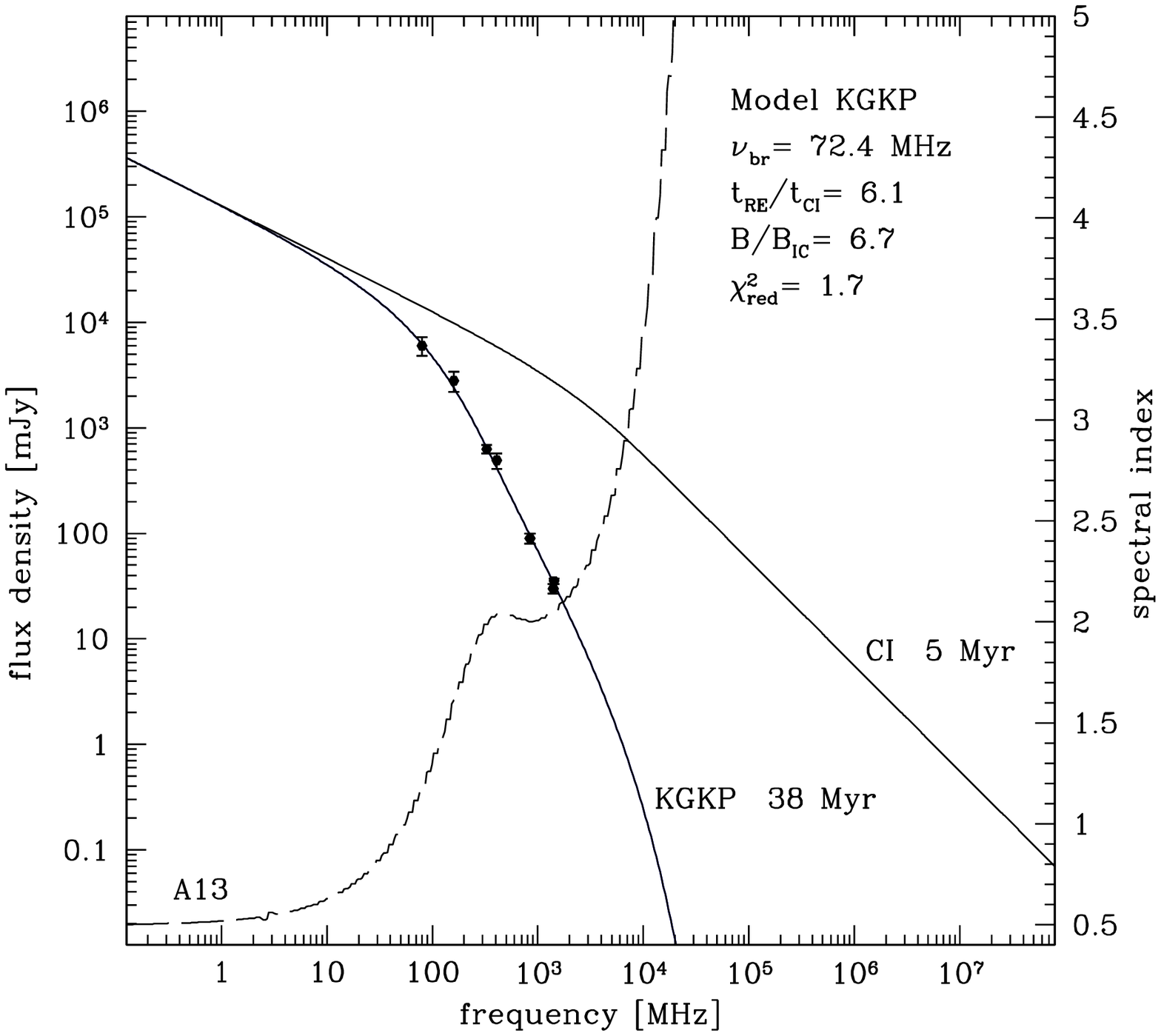,angle=0,width=8cm}
\epsfig{file=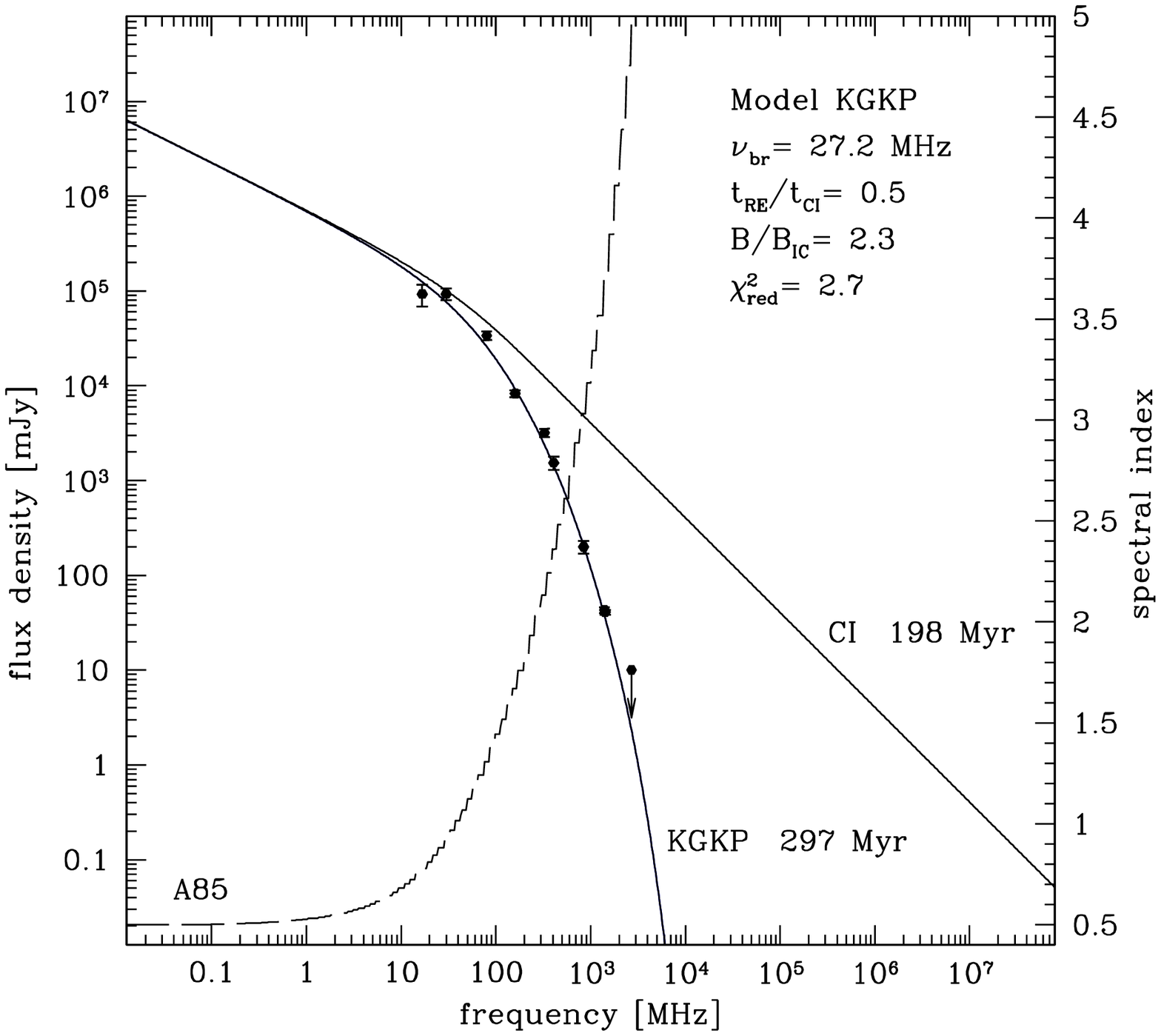,angle=0,width=8cm}
}
\mbox{
\epsfig{file=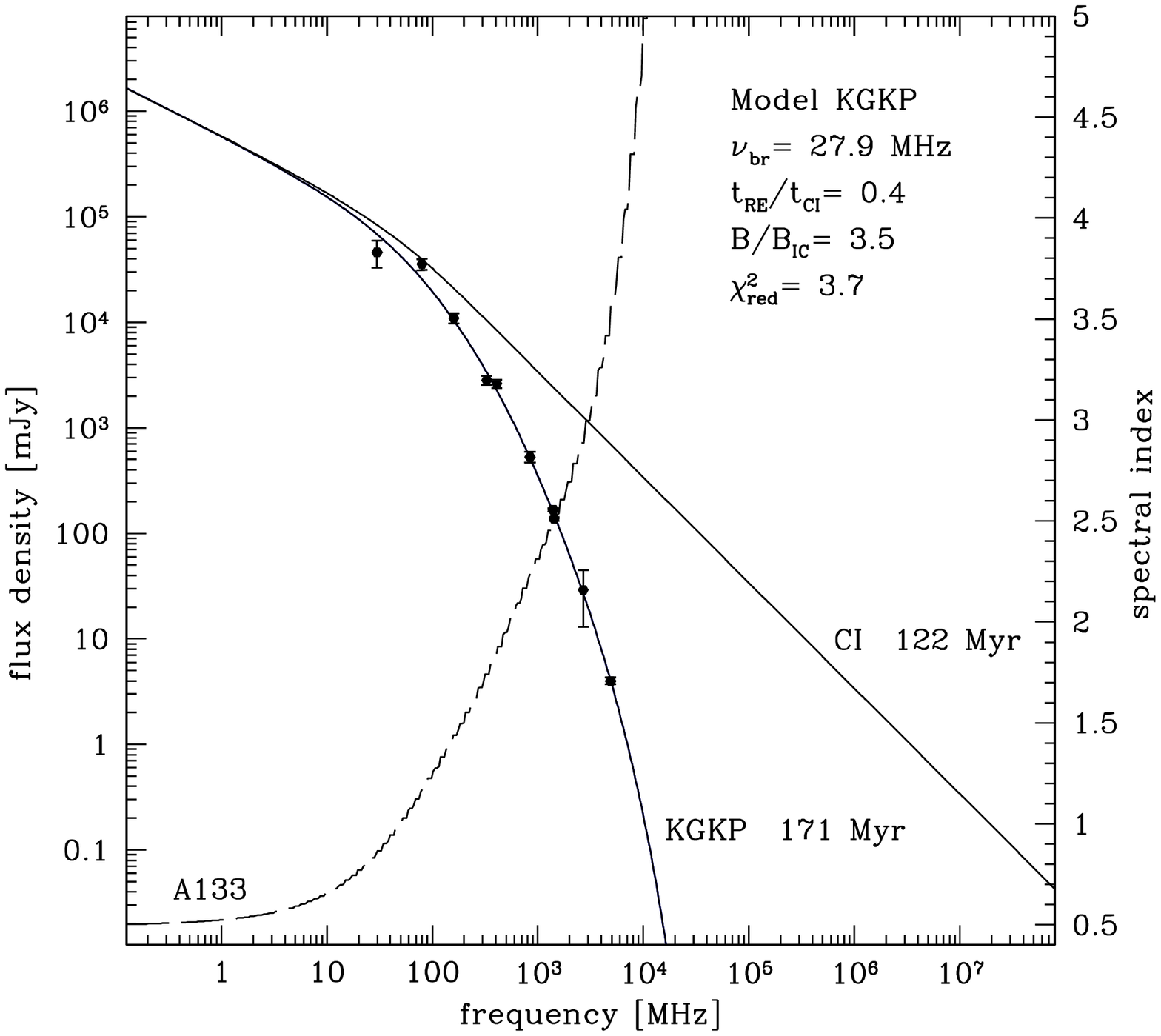,angle=0,width=8cm}
\epsfig{file=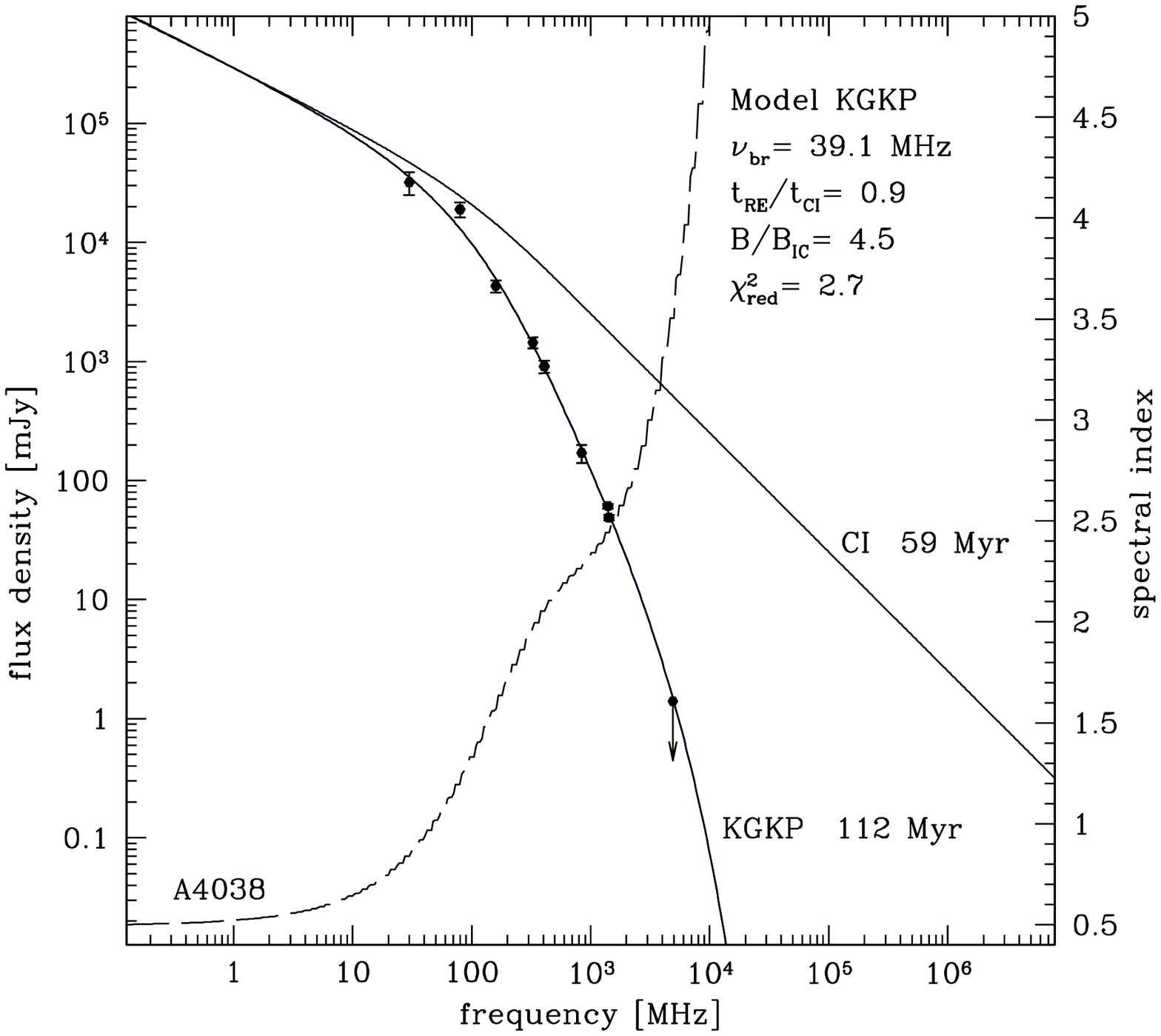,angle=0,width=8cm}
}
\caption{KGKP model fits to the spectra of four relic sources, 
based on spectral data from Table~3. The Abell name is 
indicated in the lower-left corner of each panel. 
The continuous lines are the best-fitting spectra from the
KGKP model and expected CI spectrum at the cessation of injection
(left-hand axis); the dashed line is the 
KGKP model spectral index (right-hand axis). 
Along with the model names, the CI phase duration
and the total source age are indicated.  \label{fig8}}
\end{figure}

\begin{figure}
\mbox{
\epsfig{file=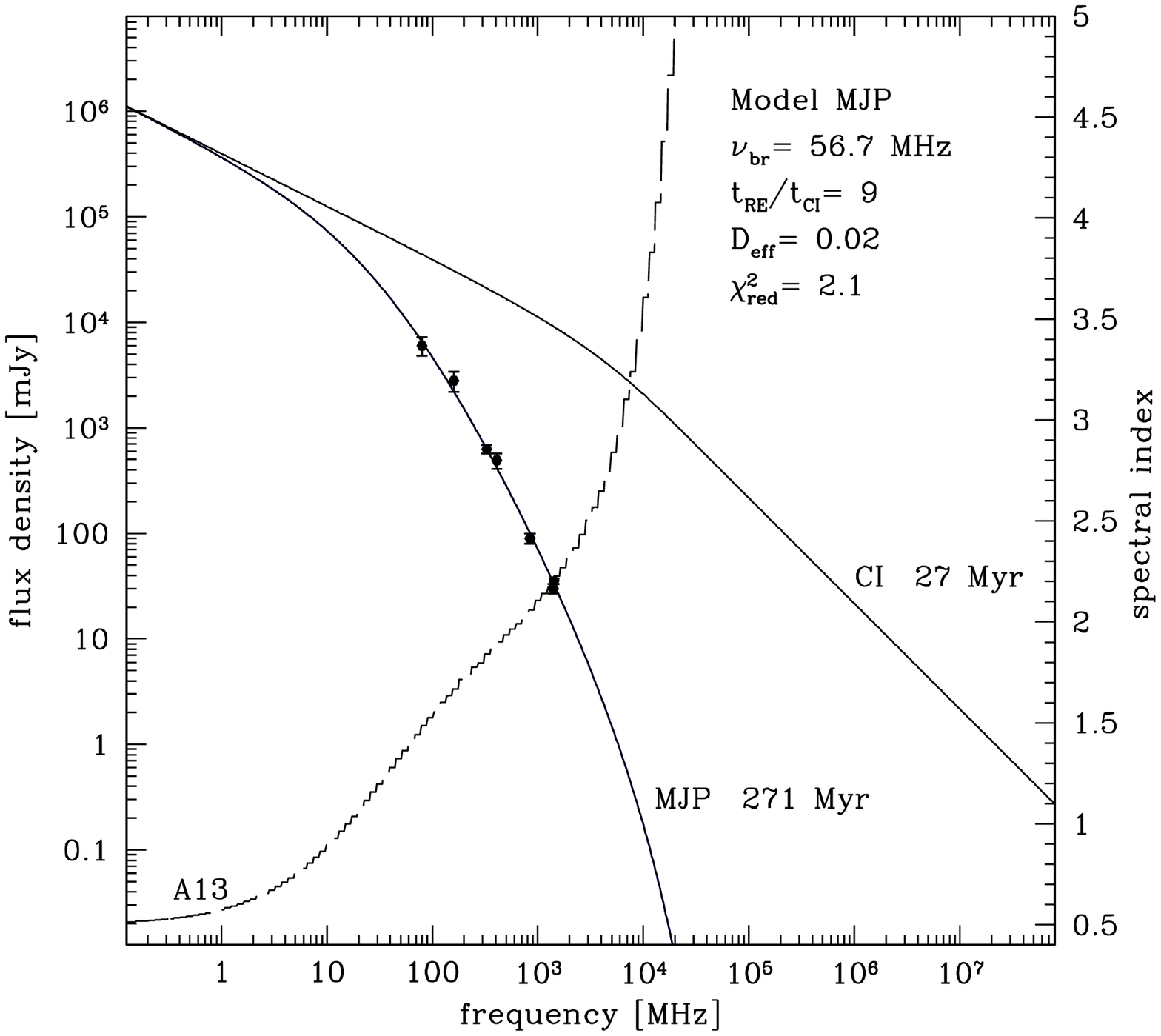,angle=0,width=8cm}
\epsfig{file=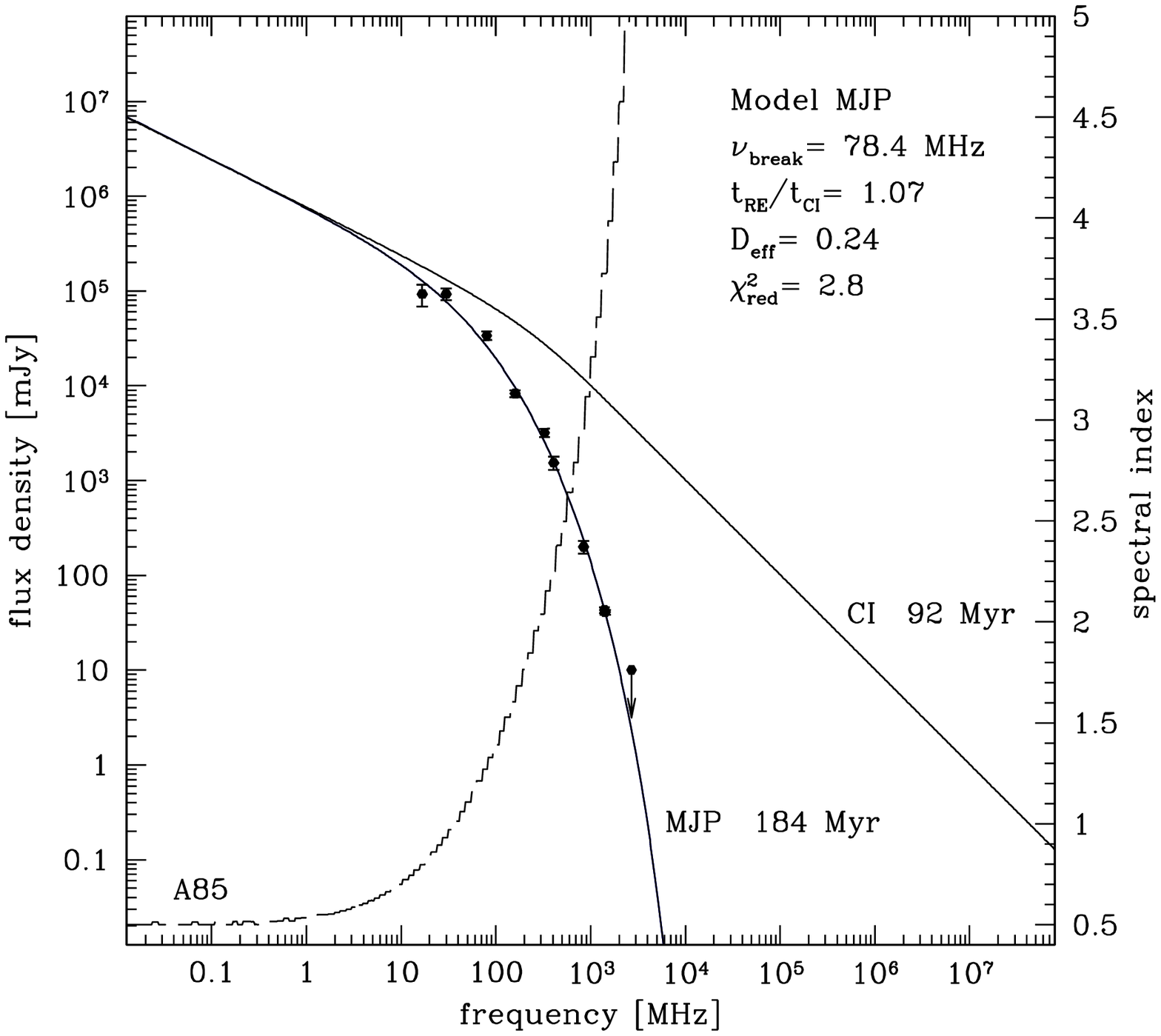,angle=0,width=8cm}
}
\mbox{
\epsfig{file=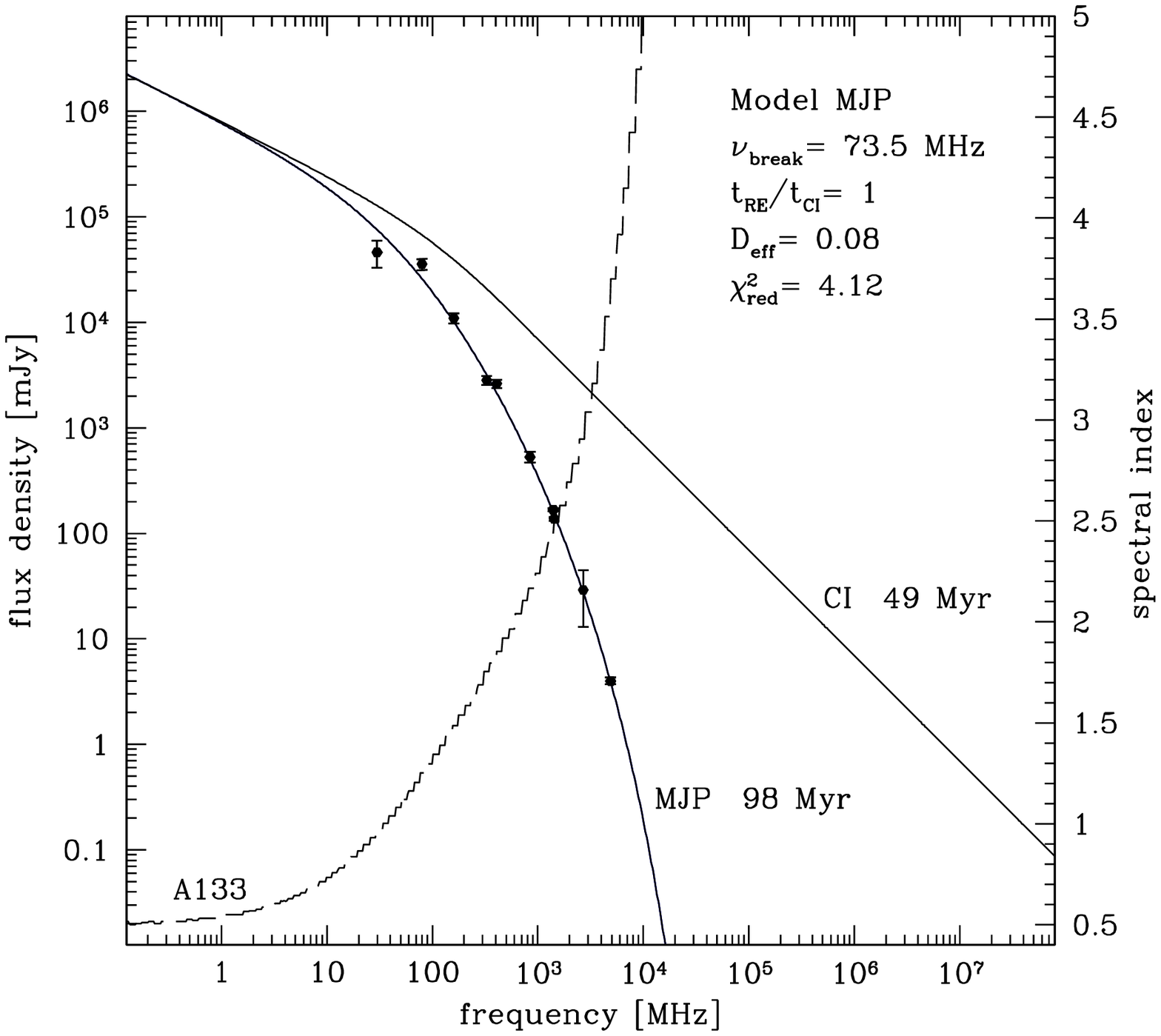,angle=0,width=8cm}
\epsfig{file=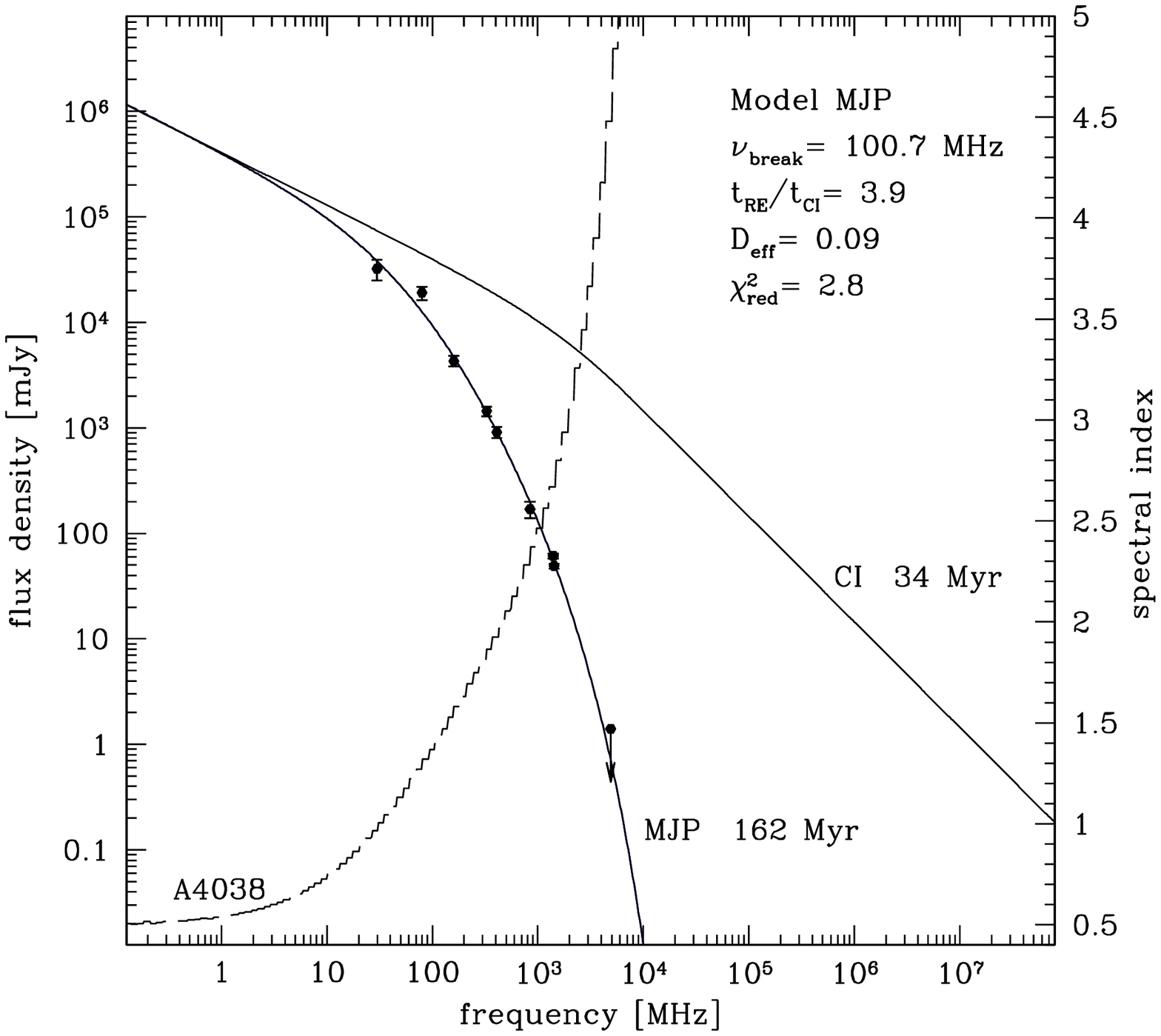,angle=0,width=8cm}
}
\caption[]{MJP model fits to the spectra of four relic sources,
based on spectral data from Table~3. The Abell name is indicated 
in the lower-left corner of each panel.  The continuous lines are
the best-fitting spectra from the MJP model and expected 
CI spectrum at the cessation of injection (left-hand axis); 
the dashed line is the MJP model spectral index (right-hand axis). 
Along with the model names, the CI phase duration and 
the total source age are indicated. \label{fig9}}
\end{figure}

\begin{figure}
\epsfig{file=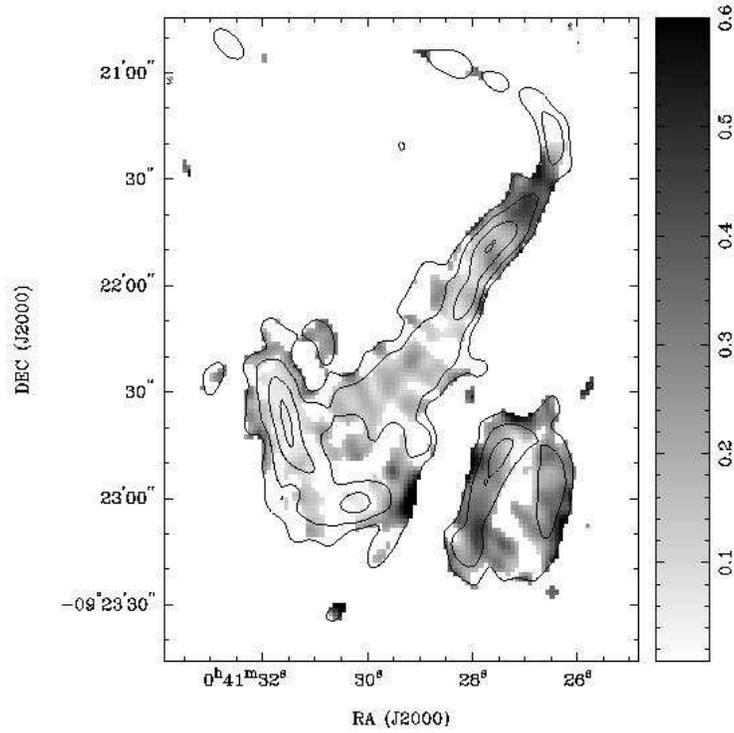,angle=-90,width=15cm,clip=true}
\caption{
Radio contours from the naturally-weighted image of the A85 relic
are overlaid on the naturally-weighted grey-scale image of 
fractional polarization at 1.425\,GHz.
Pixels are blanked when the polarized flux is $\le$\,four times 
the r.m.s.\ noise level. The polarized fraction is given by the 
vertical scale on the right-hand side. The radio contours are 
at (92, 185, 369, 738)\,$\mu$Jy\,beam$^{-1}$ or $(3.1, 6.2, 12.4, 24.7$)\,K.
\label{fig10}}
\end{figure}

\begin{figure}
\mbox{
\epsfig{file=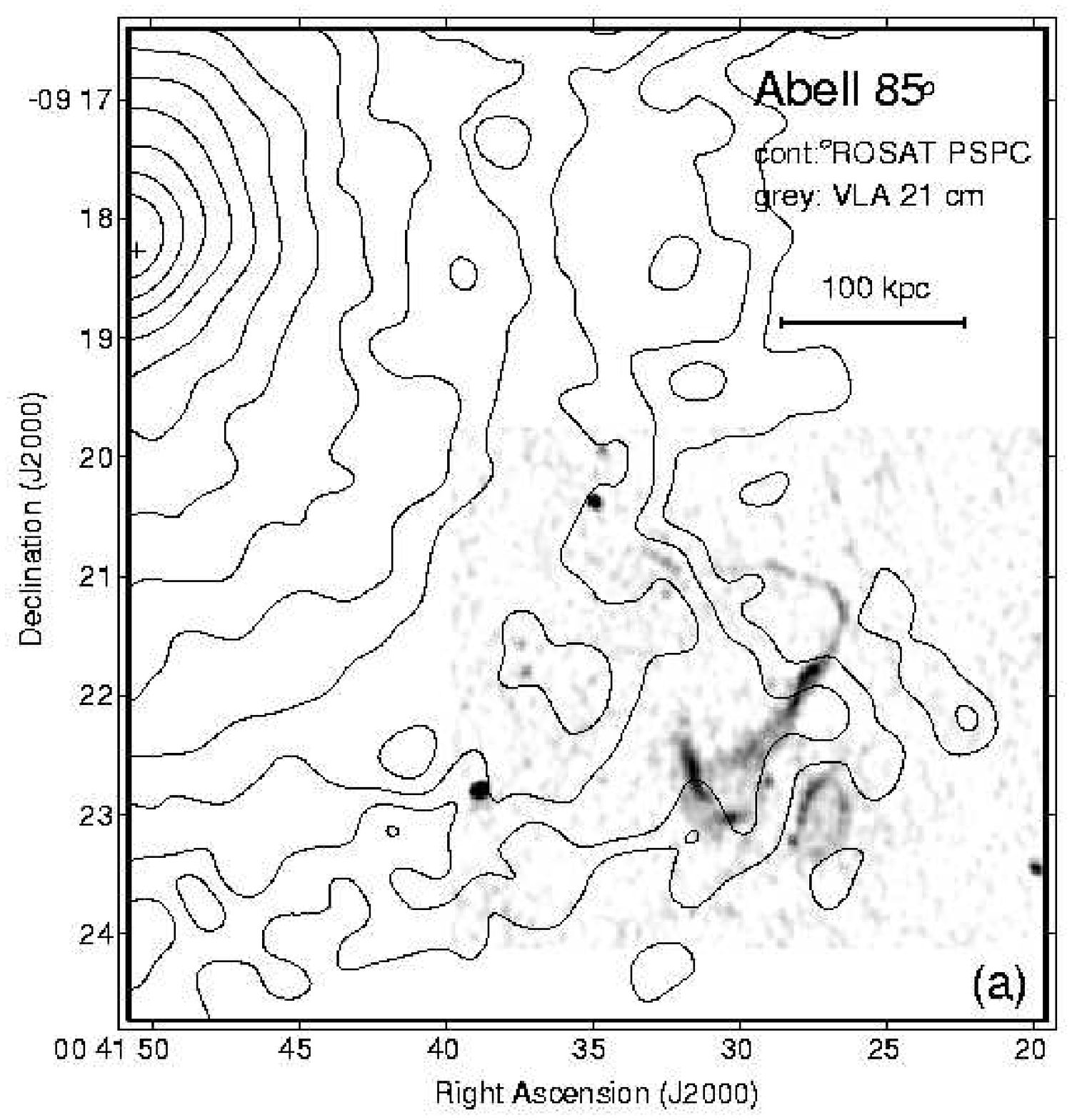,angle=0,width=8cm}
\epsfig{file=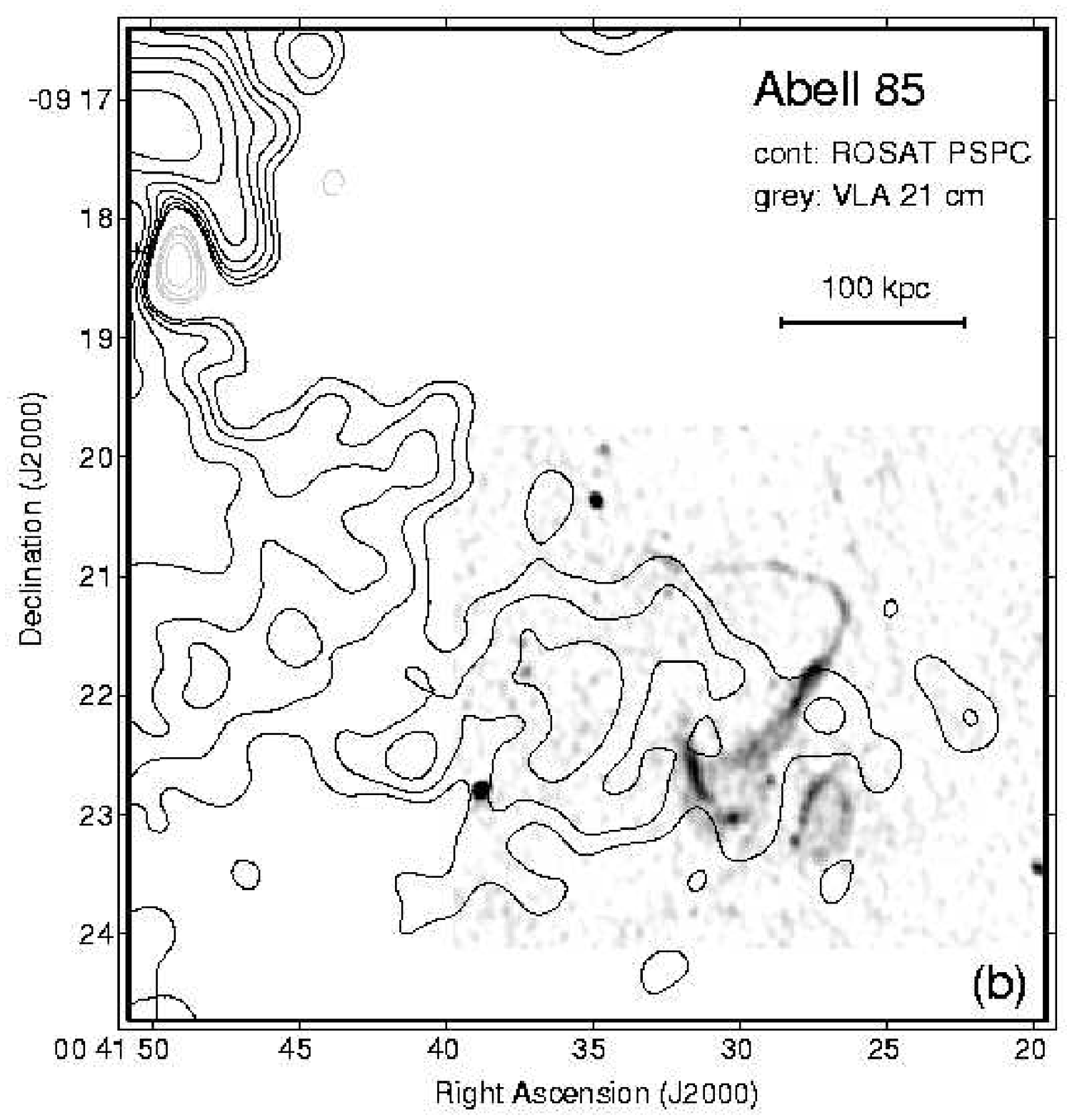,angle=0,width=8cm}
}
\caption{
South-western part of A85: (a) ROSAT PSPC hard-band (0.5\,keV to 2.0\,keV) 
contours and VLA 1.4 GHz radio image from Fig. 3 as grey-scale.
X-ray contours are at $(5, 7.07, 10, 14.1, 20, 28.3, 40, 56, 80, 113, 160, 
226, 320) \times 10^{-3}$ count s$^{-1}$ arcmin$^{-2}$ and the peak is
at $447 \times 10^{-3}$ count s$^{-1}$ arcmin$^{-2}$.  The 
resolution of the X-ray map is about 27\arcsec\,FWHM.
The position of the cD galaxy, at top left, is marked by a cross, and
the horizontal bar indicates the linear scale, based on the redshift from
Table~1.
(b) As for (a) but after subtracting a circularly-symmetric three-component
Gaussian model from the cluster X-ray emission to remove the cluster
thermal emission and improve contrast in regions of excess X-ray emission.
X-ray contours are at $(-14.1, -10, -7.07, -5, 
5, 7.07, 10, 14.1, 20, 28.3, 40)
 \times 10^{-3}$ count s$^{-1}$ arcmin$^{-2}$ with negative contours shown
in grey and the peak is at 
$52.8 \times 10^{-3}$ count s$^{-1}$ arcmin$^{-2}$.
\label{fig11}}
\end{figure}

\begin{figure}
\mbox{
\epsfig{file=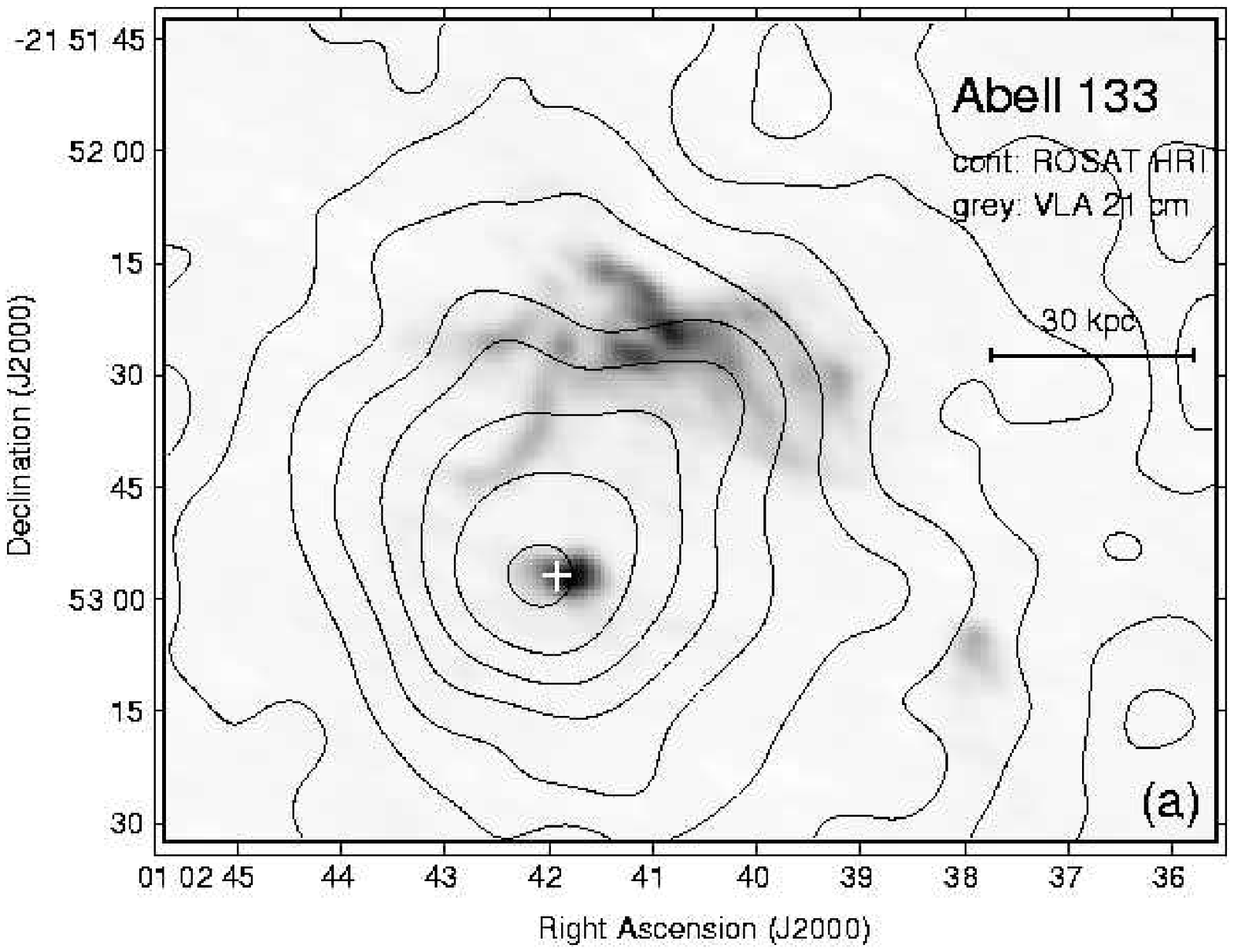,angle=-90,width=8cm}
\epsfig{file=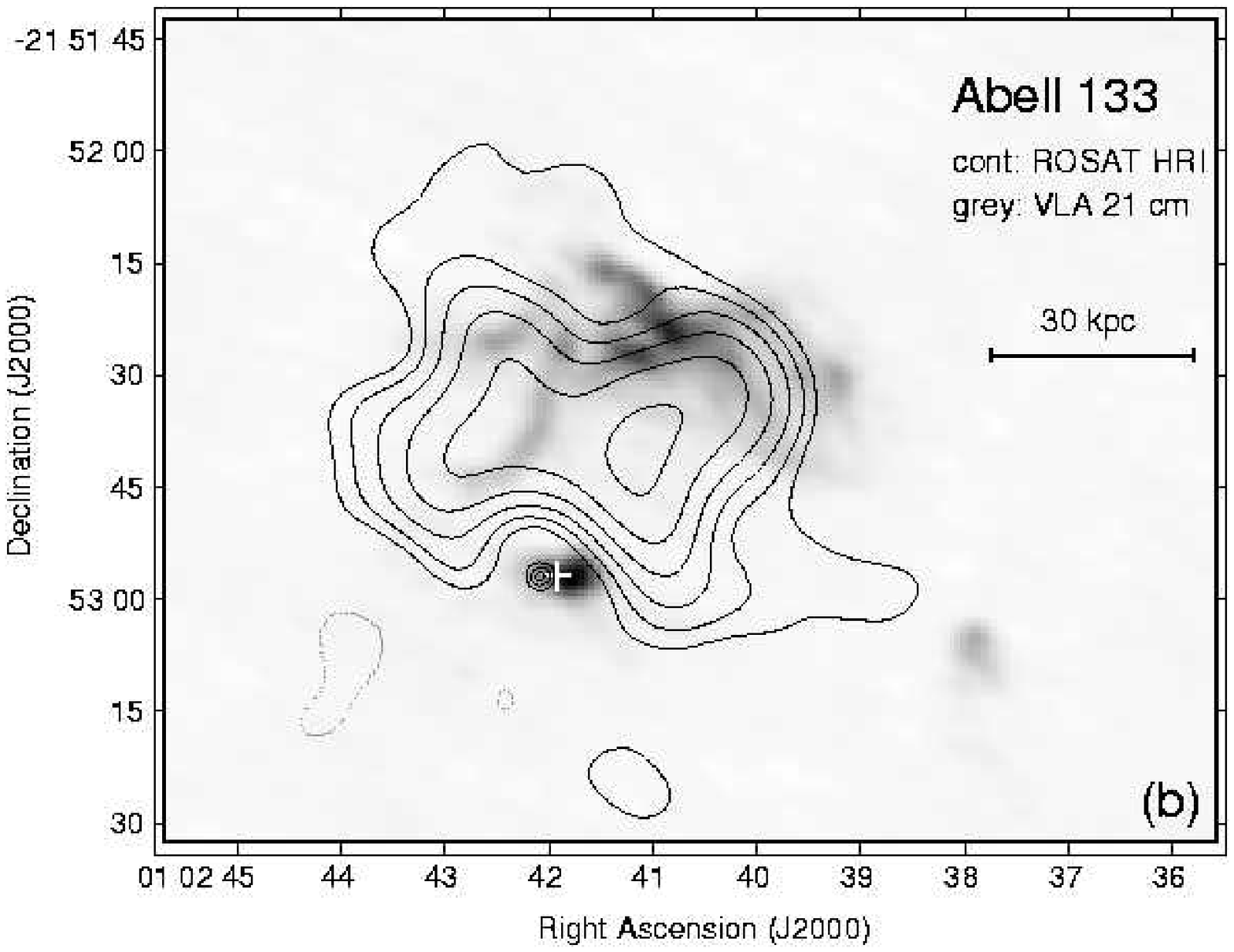,angle=-90,width=8cm}
}
\caption{
Inner part of A133: (a) ROSAT HRI (0.1\,keV to 2.4\,keV) contours and 
VLA 1.4 GHz radio image from Fig. 5 as grey-scale.  X-ray contours are at
(0.008, 0.011, 0.016, 0.023, 0.032, 0.045, 0.065, 0.091, 0.128, 0.182)
count s$^{-1}$ arcmin$^{-2}$
and the peak is at 0.193 count s$^{-1}$ arcmin$^{-2}$.
The resolution of the X-ray map is
12\arcsec\,FWHM.  The position of the cD galaxy is marked by a cross, and
the horizontal bar indicates the linear scale, based on the redshift from
Table~1.  (b) As for (a) but after subtracting an azimuthally-averaged
image of the cluster X-ray emission to remove the cluster thermal
X-ray emission and improve contrast in regions of excess X-ray
emission in the region of the relic.
X-ray contours are at (-0.008, 0.008, 0.011, 0.016, 0.023, 0.032, 0.045) 
count s$^{-1}$ arcmin$^{-2}$ with negative contours shown in grey and
the peak is at 0.0495 count s$^{-1}$ arcmin$^{-2}$. \label{fig12}}
\end{figure}

\begin{figure}
\plotone{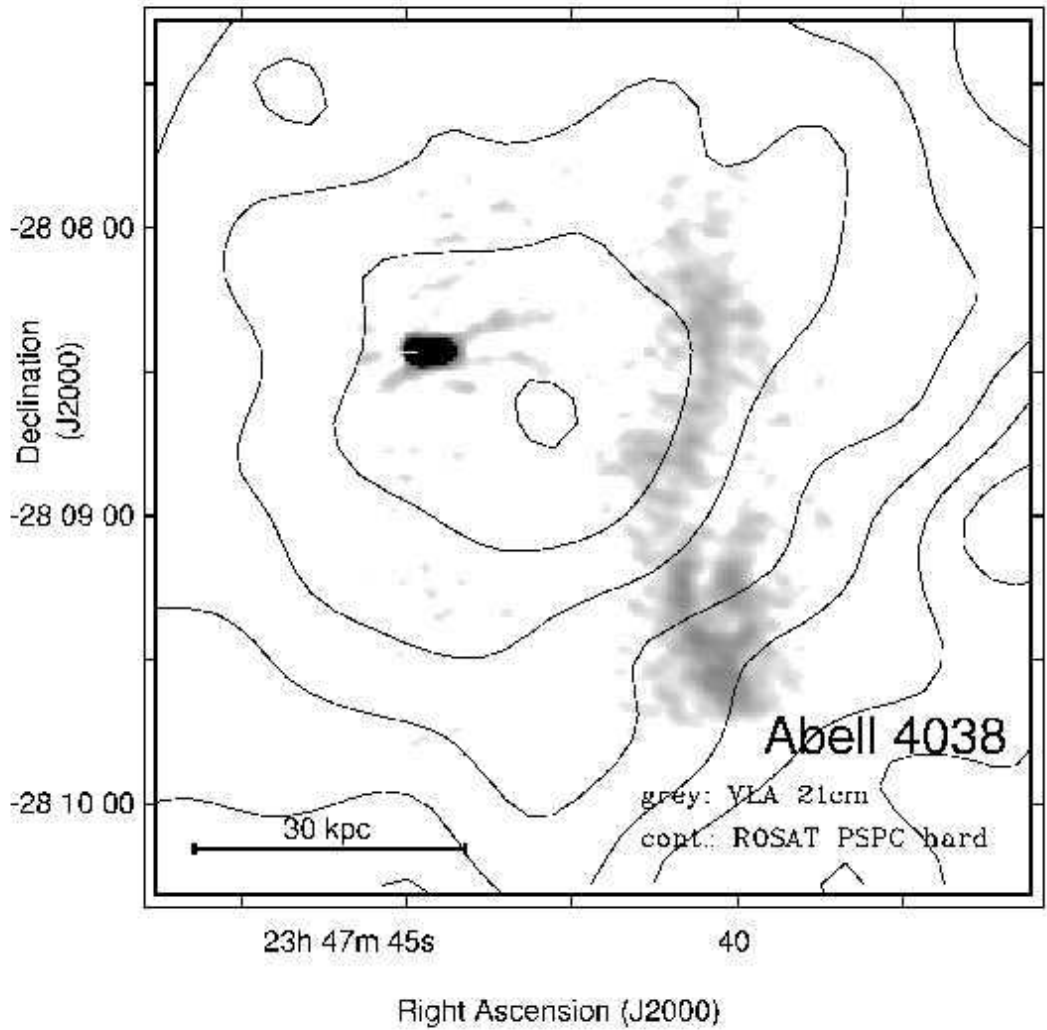}
\caption{
Inner part of A4038: ROSAT PSPC hard-band (0.5\,keV to 2.0\,keV) contours
and VLA 1.4~GHz radio image (cf.\ Slee \& Roy 1998) as grey-scale.
X-ray contours are at (5.7, 8, 16, 22.6, 32, and $45.3)\times 3.5\cdot10^{-3}$ 
count s$^{-1}$ arcmin$^{-2}$. 
The resolution of the X-ray map is about 27\arcsec\,FWHM.
The position of the cD galaxy is marked by a cross and coincides with 
the eastern end of the compact radio source.
The horizontal bar indicates the linear scale, based on the redshift
from Table~1. \label{fig13}}
\end{figure}

\end{document}